\documentclass[seceq]{ptptex}

\def\p{\partial}
\def\half{{1\over2}}
\def\real{\mathbb{R}}
\def\complex{\mathbb{C}}
\def\act{\triangleright}
\def\be{\begin{equation}}
\def\ee{\end{equation}}
\def\bea{\begin{eqnarray}}
\def\eea{\end{eqnarray}}
\def\paragraph#1{\medskip\noindent{#1}}

\def\CA{{\cal A}}   \def\CD{{\cal D}} 
 \def\CF{{\cal F}}  \def\CH{{\cal H}} 
    
 \def\CN{{\cal N}} \def\CO{{\cal O}} \def\CP{{\cal P}}

\title{
Hopf Algebra Symmetry and String Theory%
}

\author{
Tsuguhiko \textsc{Asakawa},${}^{1,}$\footnote{\tt E-mail: asakawa@nbi.dk} \ 
Masashi \textsc{Mori}${}^{2,}$\footnote{\tt E-mail: morimasa@tuhep.phys.tohoku.ac.jp}\ \ \ 
and \ 
Satoshi \textsc{Watamura}${}^{2,}$\footnote{\tt E-mail: watamura@tuhep.phys.tohoku.ac.jp} 
}

\inst{
${}^1$The Niels Bohr Institute,
The Niels Bohr International Academy,\\
Blegdamsvej 17, 2100 Copenhagen,
Denmark \\[.5em]
${}^2$Institute of Particle Theory and Cosmology\\
Department of Physics,
Graduate School of Science,\\
Tohoku University,
Sendai 980-8578, Japan 
}

\abst{
We investigate the Hopf algebra structure in string worldsheet theory 
and give a unified formulation of the quantization of string and the 
space-time symmetry.  
We reformulate the path integral quantization of string as a 
Drinfeld twist at the worldsheet level.
The coboundary relation shows that the Drinfeld twist defines a module 
algebra which is equivalent to operators with normal ordering.
Upon applying the twist, the space-time diffeomorphism is deformed 
into a twisted Hopf algebra, while the Poincar\'e symmetry is 
unchanged.
This suggests a characterization of the symmetry: 
unbroken symmetries are twist invariant Hopf subalgebras,
while broken symmetries are realized as twisted ones.
We provide arguments that relate this twisted Hopf algebra to symmetries 
in path integral quantization.
}

\begin{document}

\maketitle

\section{Introduction}
\label{sec:intro}

String theory is a promising candidate for realizing the unified 
theory of quantum gravity 
and field theory for elementary particles.
There is evidence to support the hypothesis that closed string theory 
contains general relativity.
However, it is formulated only as a perturbation theory around a 
specific background, 
and therefore the concepts in classical general 
relativity such as general covariance 
are not manifest.

As an example, consider the worldsheet theory of strings in the flat Minkowski space as a target.
The quantization of the theory should be 
Poincar\'e covariant, and thus
we find that there are massless spin 2 graviton states in the spectrum.
The massless spectrum requires, in general,  
the existence of space-time gauge symmetry and 
 diffeomorphisms, even though 
it is not a manifest symmetry of the worldsheet theory.
Moreover, scattering amplitudes among gravitons and other massless excitations are 
reproduced by the (super)gravity at low energy
(see for example Ref.\citen{Polchinski} and references therein).
It is believed that this worldsheet theory is merely an
expansion of the full string theory
(in which the diffeomorphism is manifest)
around a specific vacuum,
and the condensation of gravitons would describe another background.

From these general considerations, 
it is evident that there is a close connection
among the quantization on the worldsheet, the Poincar\'e covariance 
and the space-time gauge symmetry as well as the general covariance, 
but they are linked in a far from direct way. 
In this paper, we propose a framework to describe this connection 
by studying the Hopf algebra structure of string worldsheet theory.

The use of Hopf algebras in this paper is 
motivated by the use of the twisted Hopf algebra in the development 
of noncommutative geometry \cite{Majid,Oeckel,Sitarz0102074} and 
also by the recent progress in understanding the global or local 
symmetry on the Moyal-Weyl noncommutative space 
\cite{Watts,Oeckl2,KulishNishijima, Wess,Kobayashi}.
In Refs.\citen{KulishNishijima, Wess} and \citen{Kobayashi}
it is proposed that the explicit breaking 
of the Poincar\'e symmetry is remedied 
by considering it not as a group but as a Hopf algebra.
The key idea is that the Moyal-Weyl $*$-product is considered as a twisted product 
equipped with a Drinfeld twist of the Hopf algebra for the 
Poincar\'e-Lie algebra.
In other words, both the noncommutativity and 
the modification of the symmetry are controlled by a single twist.
It is then generalized to the twisted version of the diffeomorphism on the 
Moyal-Weyl space \cite{Wess}.
In string theory with a background $B$-field, the effective theory of 
$D$-branes is 
described by a gauge theory on the same Moyal-Weyl space \cite{SeibergWitten}.
Therefore, it is expected that there 
is a corresponding twisted Hopf algebra structure
in string theory.

However, in this paper we do not focus on the 
case with a non-zero $B$ field background; 
the purpose of this paper is 
to formulate a framework applicable to more general situations.
We will see that the twisted Hopf algebra has a similar structure 
even in a 
background with a vanishing $B$ field.
From the viewpoint of the Hopf algebra structure presented here, 
both the quantization and the space-time 
symmetry are controlled by a single twist.
Of course, we can include the nontrivial $B$-field background
in the formulation developed in this paper, 
and we will report on the case of a $B$-field background in a separate paper \cite{inpreparation}.

In this paper, we study a Hopf 
algebra structure in string worldsheet theory in the 
Minkowski background and its covariant quantization as an example, 
but in a form that enables one to apply it to more general cases.
We use the functional description of strings and define 
a Hopf algebra that consists of 
functional diffeomorphism variations as well as of
worldsheet variations,
and we also define its module algebra of classical functionals.
We then reformulate the 
path integral (functional integral) quantization of strings 
in terms of the twisted Hopf algebra for functionals.
This formulation leads to our proposal 
that each choice of twist defines a quantization scheme, 
which is a general concept not limited to our example.
By the fact that the twisted Hopf algebra is isomorphic to the original one,
but is accompanied by a normal ordering, we clarify its relation 
with the operator formulation.
Although this quantization is carried out by the twist in the 
Hopf subalgebra of worldsheet variations, the
space-time diffeomorphism is also deformed to the twisted Hopf algebra.
It turns out that the Poincar\'e-Lie algebra remains unaltered under the twist and is therefore regarded as a true symmetry, while a full diffeomorphism 
is broken but it is maintained as a twisted 
symmetry.

The paper is organized as follows:
In \S 2, we first present formulae written in standard string 
theory textbooks, 
which will be reformulated in Hopf algebra language throughout the paper.
Then, we define a Hopf algebra structure within classical string theory:
a Hopf algebra consists of functional variations including diffeomorphisms 
and a corresponding module algebra of classical functionals.
In \S 3, we
reformulate the known path integral quantization of strings 
as a twist of the Hopf algebra and the module algebra defined above.
The isomorphism between the twisted Hopf algebra 
and the normal ordered algebra is also studied 
to relate it to the operator formulation of strings.
In \S 4, we focus on the 
space-time symmetry in this twisted Hopf algebra, 
and how the twisting deforms a classical 
diffeomorphism while keeping the Poincar\'e-Lie algebra
invariant.
To relate the notion of the symmetry in the ordinary path integral, 
we rewrite Hopf algebra identities in the form of Ward-like identities 
among correlation functions.
Section 5 is devoted to a discussion and conclusion. We summarize the 
basic facts  
about Hopf algebras and their twisting in Appendix \ref{Hopf}, 
and about Hopf algebra cohomology in Appendix \ref{trivial}.
Appendices \ref{sec:cocycle-cond} and \ref{sec:Poincare} 
are devoted to technical proofs.

\section{Hopf algebra in string theory}
\label{String Hopf}

In this section, we first define the notation 
and remind the reader of some formulae in string theory that we 
will consider in this paper.
Then, we give a definition of the Hopf algebra and its module algebra 
of functionals, which appears in the classical worldsheet theory of strings.

\subsection{Preliminaries}  

We consider bosonic closed strings as well as open strings and 
take a space-filling D-brane for simplicity.
We start with the $\sigma$ model of the bosonic string 
with flat $d$-dimensional Minkowski space as the target.
The action in the conformal gauge is 
\be
S_0 [X]=\frac{1}{2\pi \alpha'}\int_{\Sigma} d^2z \eta_{\mu\nu}\p X^\mu \bar\p X^\nu ,
\label{action}
\ee
where the worldsheet $\Sigma$ 
can be any Riemann surface with boundaries, and 
typically, we take it to be the complex plane (upper half plane) 
for a closed string (open string, respectively).
$z^a=(z, \bar{z})$ are the complex coordinates on the worldsheet. 
The flat metric in the target space $\real^d$ is represented
by $\eta^{\mu\nu}$.
We frequently use the metric $\eta^{\mu\nu}$ to raise and lower the 
indices.
The worldsheet field $X^\mu (z^a)=X^\mu (z,\bar{z})$ is often  
abbreviated to $X^\mu (z)$ unless stated otherwise.

We will consider correlation functions of the form
\be
\langle V_1(z_1) \cdots V_n(z_n) \rangle _0
\label{correlation}
\ee
and their properties under space-time transformations.
Here, the vacuum expectation value (VEV) 
$\langle \cdots \rangle _0$ is defined 
by the path integral for the worldsheet field $X^\mu(z,\bar{z})$ 
with the action $S_0$ defined in (\ref{action}):
\begin{eqnarray}
\langle \CO\rangle _0=
\frac{\int \CD X \CO e^{-S_0}}{\int \CD X e^{-S_0} }~,
\label{VEV}
\end{eqnarray} 
and $V_i(z_i)$ denotes 
a vertex operator inserted at $z_i$ on the worldsheet. 
The quantization we discuss in this paper is the above path integral
average over the field $X$  under the fixed toplogy, 
since it is sufficient for describing our ideas.
Therefore, for instance, the $(b,c)$-ghost part of the full 
correlation function is omitted. We also do not perform the integration over moduli parameters since we are only considering correlation functions on the fixed worldsheet in this paper. 

In the operator formulation, 
a local vertex operator $V(z)$ is well-defined by taking 
a product of field operators $X^\mu (z,\bar{z})$, 
their derivatives 
$\p X^\mu (z)$, $\bar{\p} X^\mu (\bar{z})$, and the higher derivatives
by applying the oscillator normal ordering to avoid the divergences appearing 
in the operator product.
On the other hand, in the path integral 
there are also divergences at the coincidence point,
and these divergences are regularized either by removing the 
self-contraction by hand, or equivalently, by subtracting them
via the formula \cite{Polchinski}
\be
:\!F[X]\!:\,=\CN_0F[X]~~,\label{localoperator}
\ee
where
\be
\CN_0 = \exp\left\{-{1\over2}\int \!d^2z \!\!\int \!d^2w \,
G_0^{\mu\nu}(z,w)
\frac{\delta}{\delta X^\mu(z)}\frac{\delta}{\delta X^\nu(w)}\right\} ~,
\label{normalorderingoperator}
\ee
which is called the conformal normal ordering.
Here, $G_0^{\mu\nu}(z,w)$ is the free propagator 
on the worldsheet defined through
\be 
\langle X^\mu(z)X^\nu(w)\rangle_0=G_0^{\mu\nu}(z,w)\ 
\label{propagator1}
\ee
and its function form depends on the worldsheet topology.
For instance, on the complex plane, it is
\be
G_0^{\mu\nu}(z,w):=\eta^{\mu\nu}G_0(z,w)=-{\alpha' \over 2}\eta^{\mu\nu} \ln |z-w|^2.
\label{tree propagator}
\ee
In this case, the conformal normal ordering coincides with the oscillator normal ordering,
but the subtraction using (\ref{tree propagator}) works in general and 
we refer to (\ref{localoperator}) as the normal ordering in the following.
Note that $:\!F[X]\!:$ itself is a power series expansion 
with divergent coefficients, 
and therefore it should be understood only together with the path integral.

The local vertex operators may be 
located either in the bulk or on the boundary of the 
worldsheet.
The product of two local vertex operators in any correlation function
is a time-ordered product in the operator formalism,
which has the natural correspondence with the path integral formulation.
It is rewritten by the normal ordered (but bi-local) vertex operators using 
Wick's theorem:
\bea
\label{Pol-identity}
&&:\!F[X]\!:\, :\!G[X]\!: \nonumber \\
&=&:\!\exp\left\{\int \!d^2z \!\!\int \!d^2w \,
\eta^{\mu\nu}G_0(z,w)
{\delta_F\over\delta X^\mu(z)}{\delta_G\over\delta X^\nu(w)}\right\} 
F[X]G[X]\!:,
\eea
where the derivative $\delta_F (\delta_G) $ acts on $F$ ($G$). 
This formula is again valid only inside
the path integral VEV. 
In addition, by using the Taylor expansion around $w$ with 
respect to $(z-w)$, 
the r.h.s. coincides with the usual operator product expansion.
As we shall see, the above formulae (\ref{normalorderingoperator}) 
and (\ref{Pol-identity}) 
always require careful application at the functional level.
One of the purposes of this paper is to give a simple algebraic 
characterization of 
these formulae as Hopf algebra actions.
In this formulation, not the operators 
but only the functional calculi are used, and 
there are no complications when the formal divergent 
expansions appear in the normal ordering formula.

After formulating the VEV
in functional language, we shall discuss the symmetry of the 
correlation functions  (\ref{correlation}).
If the action $S_0$ and the measure are invariant under the variation of
the worldsheet fields 
$X^\mu(z) \rightarrow X^\mu(z) +\delta X^\mu(z)$,
this defines a symmetry in the quantum theory, and we obtain a
 Ward identity associated with such a variation:
\be
\label{Ward id}
0= \sum_{i=1}^{n} \langle V_1(z_1) \cdots \delta V_i(z_i) \cdots V_n(z_n) \rangle _0.
\ee
Here the infinitesimal transformation of any local vertex operator 
$V(z)=:\!F[X]\!:(z)$
is given by the commutation relation with the symmetry generator,
and it has the same form as the classical transformation.
It is written in terms of first-order functional derivatives as
\be
\delta V(z) =-:\! \int \! d^2w \,\delta X^\mu(w)
\frac{\delta F[X]}{\delta X^{\mu}(w)}\!:~~.
\label{deltaF}
\ee
In our case, the unbroken space-time symmetry consists of 
Poincar\'e transformations generated by
\begin{eqnarray}
&&P^{\mu}=-i \int d^2z \,\eta^{\mu\lambda}
\frac{\delta}{\delta X^{\lambda}(z)},\nonumber \\ &&
L^{\mu\nu}=-i \int d^2z  X^{[\mu}(z) \eta^{\nu]\lambda} 
\frac{\delta}{\delta X^{\lambda}(z)}, 
\label{generators}
\end{eqnarray}
where $P^\mu$ are the generators of the translation and $L^{\mu\nu}$ are 
the Lorentz generators.
Note the position of the normal ordering operation in (\ref{deltaF}).
To obtain the quantum transformation law and a similar identity 
in the case that a variation is not a symmetry, 
we again need to take care of the ordering and the 
divergences.
We see in \S \ref{sec:symmetry} that the transformation law of 
broken symmetries
should be twisted in the Hopf algebra sense.

\subsection{Hopf algebra for classical functional variations}
\label{sec:classical}

Before discussing the quantized theory of strings,
we consider a Hopf algebra structure and the related 
module algebra structure at 
the classical level, which underlies 
the quantization of the string worldsheet theory.
We will use functionals and functional derivatives as our main tools.
Actually, this structure does not depend on the 
action $S_0$, nor on the conformal symmetry,
and thus it is background independent.


\paragraph{\it Classical functionals as module algebra}

Classically, the string variable $X^\mu(z)\,(\mu=0,\cdots d-1)$ is a set of classical 
functions defining the
embedding map $X$ of a worldsheet $\Sigma$ into a target space $\real^d$:
\be
X: \Sigma\ni z \mapsto 
X(z)=\left( X^0(z,\bar z),\cdots, X^{d-1}(z,\bar z)\right) \in \real^d 
\ee 
Any function on the space-time $\real^d$ is mapped to the worldsheet function 
via the pull-back $X^* :C^\infty(\real^d) \rightarrow C^\infty(\Sigma)$ as
$f \mapsto (X^* f)(z)=f[X(z)]$.
The pull-back of a $1$-form
$\omega \in \Omega^1(\real^d)$ is also defined by
$X^* (\omega_\mu dX^\mu) = \omega_\mu [X(z)] \p_a X^\mu(z) dz^a \in\Omega^1(\Sigma)$.
They can be extended to any tensor field on the space-time. 
Therefore, any field on the space-time (D-brane) is realized as a worldsheet field.
For example, a scalar (tachyon) field and a gauge field give
\bea
(X^* \phi)(z)&=&\phi[X(z)] =\int \! d^d k \phi(k) e^{ikX(z)}~,\nonumber\\
(X^* A_\mu)_a (z)& =& A_\mu[X(z)] \p_a X^\mu(z)~.
\label{tensor pull-back}
\eea

A complex valued functional $I[X]$ of $X$ is defined on the space of embeddings
as a $\complex$-linear map $I:{\rm Map}(\Sigma,\real^d)\rightarrow\complex$.
It is typically given by the integrated form over the world sheet $\Sigma$ as
\be
I[X]= \int \!d^2 z\ \rho(z) F[X(z)]~ .
\label{def of functional}
\ee
where $F[X(z)]$ is a component of a pull-back tensor field such as 
(\ref{tensor pull-back}) 
and $\rho(z)$ is a weight function (distribution).
The action functional $S_0[X]$ in (\ref{action}) is a simple example.
Note that a pull-back function $F[X(z)]$ defines a functional 
when we fix $z$ at some point $z_i \in \Sigma$.
Thus, we also consider a functional with an additional label $z_i$ by choosing the delta function 
as the weight function $\rho(z)$,
\be
F[X](z_i)= \int \!d^2 z\ \delta^{(2)}(z-z_i) F[X(z)]~ ,
\label{local functional}
\ee
which we call a local functional at $z_i$.   
We also write it simply as $F[X(z_i)]$ when this does not cause confusion. 
The types of functionals given by (\ref{def of functional}) and 
(\ref{local functional}) 
correspond to an integrated vertex operator 
and a local vertex operator after quantization, respectively.

Now let $\CA$ be the space of complex valued functionals 
comprising of the embedding $X^\mu(z)$ 
and its worldsheet derivatives $\p_a X^\mu(z)$ described above. 
We define the multiplication of two functionals as
$I_1I_2[X]=I_1[X]I_2[X]$, where the r.h.s. is multiplication 
in $\complex$.
This leads to the multiplication of two local functionals as
$FG[X](z_1,z_2)=F[X(z_1)]G[X(z_2)]$.
In order that this product is an element of $\CA$, bilocal 
functionals at $(z_1,z_2)$ should be 
included in $\CA$.
By including all multi-local functionals with countable labels, $\CA$ forms 
an algebra over $\complex$. 
We denote this product as a map $m : \CA\otimes \CA \rightarrow \CA$:
\be
m(F\otimes G)=FG~.
\ee
Note that the product is commutative and associative.

\paragraph{\it Hopf algebra of functional vector fields}

Next, let us define a Hopf algebra acting on the classical 
functionals $\CA$.
Consider an infinitesimal variation of the embedding function
$X^\mu(z) \rightarrow X^\mu(z) +\xi^\mu[X(z)]$, 
which is a diffeomorphism from the viewpoint of the target space.
Then, the change of a functional is 
generated by a first order functional derivative of the form
\be
\label{vector field}
\xi = \int \! d^2w \, \xi^\mu[X(w)] \frac{\delta}{\delta X^{\mu}(w)},
\ee
where the functional derivative is defined by
\be
{\delta\over\delta X^\mu(z)} X^\nu (w):=\delta^\nu_\mu \delta^{(2)}(z-w)~.
\ee
The object $\xi$ in (\ref{vector field}) is a functional version 
of the vector field acting on $\CA$
and it is a derivation of the algebra $\CA$.
For a local functional $F[X]$ its action (Lie derivative along $\xi$) is 
written as $\xi \,\act F[X]= (\xi^\mu \p_\mu F)[X]$.
It is related to the variation of the functional under the diffeomorphism as
$\delta_\xi F[X] = -\xi\act F[X],$\
\footnote{
Note that $F[X]$ is a scalar functional so that $F'[X']=F[X]$.  
As usual, the variation of $F[X]$ is defined by the difference at the same ``point" $X$ 
and it is written by ($-1$ times) the Lie derivative along $\xi$ as
$\delta_\xi F[X] =F'[X]-F[X]=-(\xi^\mu \p_\mu F)[X]$.
}

The object $\xi$ can be extended to the following expression 
by including world sheet variations: 
\be
\label{gene vector field}
\xi = \int \! d^2w \, \xi^\mu (w) \frac{\delta}{\delta X^{\mu}(w)},
\ee
where $\xi^\mu (w)$ is a weight function 
(distribution) on the worldsheet 
of the following two classes.
\begin{itemize}
\item[i)] $\xi^\mu(w)$ is a  pull-back of a target space function $\xi^\mu (w)= \xi^\mu [X(w)]$.
It corresponds to a target space vector field as defined above.
\item [ii)] $\xi^\mu (w)$ is a function of $w$ but is independent of $X(w)$ and its derivatives.
It corresponds to a change of the embedding 
$X^\mu(z) \rightarrow X^\mu(z) +\xi^\mu(z)$ and is
used to derive the equation of motion.
We also admit functions such as $\xi^\mu(w,z_1,\cdots)$ with some additional labels 
$z_1, \cdots$.
The functional derivative itself is an example of this, i.e., by setting 
$\xi^\mu(w,z)=\delta^\mu_\nu \delta (w-z)$ in eq. (\ref{gene vector field}).

\end{itemize}
Such a mixture of space-time vector 
fields and worldsheet variations becomes  
important in the following sections.
Note that in this paper we do not consider another class with
 $\xi^\mu(w)=\epsilon^a (w) \p_a X^\mu(w)$, corresponding to an
infinitesimal coordinate transformation 
$w^a \mapsto w^a +\epsilon^a (w)$ on the 
worldsheet.

We denote the space of all such vector fields $\xi$ (\ref{gene vector field}) 
as ${\mathfrak X}$ 
and, in particular, the $\xi$ in class ii) as ${\mathfrak C}$.
We write its action on $\CA$ as $\xi \triangleright F$.
By successive transformations 
$\xi \triangleright (\eta \triangleright F)$, 
we see that functional vector fields form a Lie algebra with the Lie bracket
\be
\label{Lie algebra}
[\xi, \eta]= \int \! d^2w \, \left(
\xi^\mu \frac{\delta \eta^\nu}{\delta X^{\mu}}
-\eta^\mu \frac{\delta \xi^\nu}{\delta X^{\mu}}
\right) (w) \frac{\delta}{\delta X^{\nu}(w)}~~.
\ee
We can then define the universal 
enveloping algebra $\CH=U({\mathfrak X})$ of ${\mathfrak X}$ over $\complex$,
which has a natural cocommutative Hopf algebra structure
$(U({\mathfrak X}); \mu,\iota,\Delta,\epsilon,S)$
\footnote{This is a generalization of the Hopf algebra of vector field discussed in Ref.\citen{Wess}. 
For a similar approach see also Ref.\citen{Aschieri}.}. 
The defining maps given on elements $\xi, \eta \in {\mathfrak X}$ are
\begin{eqnarray}
&& \mu (\xi \otimes \eta )=\xi \cdot \eta~ , ~~~~ 
\iota(k)=k\cdot 1~, \nonumber \\ 
&&\Delta(1)=1\otimes 1~, ~~~\Delta(\xi )=\xi  \otimes 1+1 \otimes \xi~,  \nonumber \\ 
&&\epsilon(1)=1~,  ~~~\epsilon(\xi)=0~, \nonumber \\ 
&&S(1)=1, ~~~S(\xi )=-\xi~, 
\end{eqnarray}
where $k\in\complex$.
The copropduct $\Delta(\xi ) $ implies that $\xi$ is a primitive element, 
which follows from the Leibniz rule of functional derivatives.
The product $\mu$ is defined by successive transformations 
of $\eta $ and $\xi$ and it is also denoted by $\xi \cdot \eta $.
It gives higher order functional derivatives 
thus that the vector space $U({\mathfrak X})$ 
consists of elements of the form 
\be
h = \int \! d^2 z_1 \cdots \int \! d^2 z_k \, \xi^{\lambda_1} 
(z_1)
\frac{\delta}{\delta X^{\lambda_1}(z_1)} \cdots 
\xi^{\lambda_k} (z_k)
\frac{\delta}{\delta X^{\lambda_k}(z_k)}~~.
\ee
As usual, the maps are uniquely extended to any such element of $U({\mathfrak X})$ by the algebra (anti-) homomorphism.

The algebra $\CA$ of functionals is now considered to be an 
$\CH$-module algebra.
The action of the element $h \in \CH$ on $F \in \CA$ is denoted by
$h \triangleright F$ as above. 
The action on the product of two elements in $\CA$ is defined by 
\be
h \triangleright m (F\otimes G) = m \Delta(h) \triangleright (F\otimes G)~,
\label{covariance}
\ee 
which represents the covariance of the module algebra $\CA$ under 
diffeomorphisms 
or worldsheet variations.

In particular, the Poincar\'e transformations 
are generated by (\ref{generators}).
It is easy to see that they satisfy the standard commutation relation for 
the Poincar\'e-Lie algebra, $\CP=\real^d \ltimes {\mathfrak so}(1,d-1)$, 
and that $\CP$ is a Lie subalgebra of ${\mathfrak X}$.
As a result, their universal envelope 
$U(\CP)$ is also a Hopf subalgebra of $\CH=U({\mathfrak X})$.

Another Hopf subalgebra $U({\mathfrak C}) \subset \CH$ 
is that generated by worldsheet variations.
Such variations form an abelian Lie subalgebra ${\mathfrak C}$, 
and thus the algebra $U({\mathfrak C})$ is a commutative and cocommutative Hopf algebra.

\section{Quantization as a twist on the worldsheet}
\label{sec:quantization}

So far we have only dealt with {\it classical} functionals of $X^\mu (z)$.
In string theory, the field $X^\mu (z)$ must be quantized. 
The quantization can be achieved using the functional integral over 
all possible embedding functions $\{ X^\mu (z)\}$ weighted with a 
Gaussian-type 
functional $e^{-S_0 [X]}$ as given in eq. (\ref{VEV}).
Since $S_0$ is quadratic in $X$, the VEV of a functional 
$\langle \,I[X]\, \rangle_0$ 
is completely determined by the Wick contraction.
We show that the same VEV is reproduced simply by a twist of a Hopf algebra.
This leads to our proposal that a Hopf algebra twist is a quantization.
By using cohomological results, we clarify the relation between 
the twist and the normal ordering, which gives a more rigorous 
characterization of the path integral VEV.

\subsection{Wick contraction as a Hopf algebra action}

We first take a heuristic approach to rewriting the path integral VEVs 
 in terms of the Hopf algebra $\CH$ introduced in \S \ref{sec:classical}.
To this end, we consider the following two maps 
$\CN_0^{-1} \, \act : \CA\rightarrow \CA$ and $\tau:\CA\rightarrow \complex$ where
\bea
&&\CN_0^{-1}=\exp\left\{{1\over2}\int \!d^2z \!\!\int \!d^2w \,
G_0^{\mu\nu}(z,w)
{\delta\over\delta X^\mu(z)}{\delta\over\delta X^\nu(w)}\right\}
\label{N_0} ~, \\
&&\tau (I[X])=I[X]\Big|_{X=0}~~.
\label{def of tau}
\eea
Here $\CN_0^{-1}$ 
is an element of $\CH$ and this Hopf algebra action 
gives the contraction with respect to the free propagator (\ref{propagator1}), while 
$\tau$ extracts the scalar terms 
independent of $X$ in the functional. 
Note that if $I[X]$ contains a local functional $F[X](z)$, 
$\tau(I[X])$ also depends on the label $z$ in general, 
i.e., it is a complex function of $z$.

The path integral average (\ref{VEV}) 
of a functional $I[X] \in \CA$ can be written 
as a composition of these maps as 
$\tau\circ\CN_0^{-1}\,\act :\CA \rightarrow \complex$:
\be
\langle \,I[X]\, \rangle_0
= \tau  (\CN_0^{-1}\,\act I[X])~.  \label{VEVpathintegral}
\ee
This is simply a rewriting of the formula 
derived in the standard path integral argument\cite{Polchinski}:
We consider the generating functional $Z[J]$ 
by temporarily introducing the source $J_\mu(z)$ for $X^\mu (z)$, 
where the VEV of $I[X]$ is given as $I[{\delta \over \delta J}]Z[J]|_{J=0}$.
Then, upon removing $J_\mu(z)$ from the expression by replacing it 
with the functional derivative of $X^\mu$, we obtain 
(\ref{VEVpathintegral}).
Therefore, we also call \eqref{VEVpathintegral}  
the VEV as in the case of the path integral.

However, for any functional $I[X]$ corresponding to a composite 
operator of $X$, 
the above map suffers from divergences originating from self-contractions. 
To remove these divergences, each functional inserted into the VEV is considered to be
a normal ordered functional.
Let $I[X]$ be a single local functional $F[X](z)$.  
The normal ordering is also given by a Hopf algebra action:
\be
:\!F[X]\!:(z)=\CN_0 \triangleright F[X](z)~, 
\ee
where the subtraction $\CN_0 \in \CH$ 
is given in eq. (\ref{normalorderingoperator}), 
which is the inverse of the contraction $\CN_0^{-1}$ (\ref{N_0}) in $\CH$. 
Then, its VEV is
\be
\langle \,:\!F[X]\!:(z) \,\rangle_0
= \tau(F[X](z))~~.
\label{def of vacuum}
\ee
In particular, for any normal ordered local functional without a scalar term,
its VEV is zero.
It is known that this (conformal) normal ordering 
coincides with the oscillator normal ordering for $\Sigma=\complex$.
In that case, (\ref{def of vacuum}) corresponds to the characterization of the oscillator 
vacuum.

If the functional $I[X]$ is 
a multi-local functional at $z_1, z_2,\cdots$ 
given by a product of these normal ordered functionals, 
the path integral formula is given exactly by 
eq. (\ref{VEVpathintegral}) which leads to the 
multi-variable functions of $z_1, z_2,\cdots$.
In particular, let us consider the VEV of the product of 
two local functionals $:\!F[X]\!:\!(z)$ and $:\!G[X]\!:\!(w)$, given by
\be
\sigma(z,w)=\langle\, :\!F[X]\!:\!(z) :\!G[X]\!:\!(w) \,\rangle_0~.
\label{<FG>}
\ee
Using the above introduced maps, we can rewrite the correlation function
as a sequence of maps
as
\bea
\sigma(z,w)
&=&\tau \circ \CN_0^{-1} \act m \left[
(\CN_0 \otimes \CN_0) \act (F[X] \otimes G[X])
\right] \nonumber \\
&=&\tau \circ m \left[\Delta (\CN_0^{-1} )
(\CN_0 \otimes \CN_0) \act (F[X] \otimes G[X])
\right]~,\label{twocorrelation}
\eea
where in the second line we used the covariance (\ref{covariance}) 
of a Hopf algebra action on 
the product.
This coproduct $\Delta (\CN_0^{-1}) \in \CH\otimes \CH$ 
shows that Wick contractions act separately on both 
$F$ and $G$ as self-contractions, and also 
as intercontractions between $F$ and $G$, 
but, because of the $(\CN_0 \otimes \CN_0)$ factor, 
only the latter is effective.
Thus, the net contraction 
is characterized by an element of $\CH\otimes \CH$,
\be
\CF_0^{-1}=\Delta(\CN_0^{-1})(\CN_0\otimes\CN_0)~.
\label{above relation}
\ee
We show that the inverse of this operator defined by 
\be
\CF_0:=\exp\left\{-\int \! d^2z \!\! \int \! d^2w \,
G_0^{\mu\nu}(z,w){\delta\over\delta X^\mu(z)}
\otimes {\delta\over\delta X^\nu(w)}\right\}~~.
\label{F_0}
\ee
satisfies (\ref{above relation}).
For this, we write $\CF_0=\exp(F_0)$ in (\ref{F_0}) and $\CN_0=\exp(N_0)$ 
in (\ref{normalorderingoperator}).
Using the explicit form of $N_0$, the coproduct of $N_0$ is given by 
\be
\Delta (N_0)=N_0 \otimes 1+1\otimes N_0 + F_0~~.\label{exponentrel}
\ee
from the standard Leibniz rule of the functional derivative. 
Here we have used the fact that $F_0$ is symmetric under the exchange of tensor factors, owing to the property of the Green function: $G_0^{\mu\nu}(z,w)=G_0^{\nu\mu}(w,z)$.
Then, the relation (\ref{exponentrel}) leads to
\bea
\Delta (\CN_0) &=& \Delta (e^{N_0}) =e^{\Delta (N_0)}=e^{N_0\otimes 1+1\otimes N_0+ F_0}
\nonumber\\
&=& (\CN_0\otimes 1)(1\otimes \CN_0)\CF_0
=(\CN_0\otimes \CN_0)\CF_0~.
\eea
Therefore, $\CF_0$ (and $\CF_0^{-1}$) is written in terms of $\CN_0$ as follows:
\bea
&&\CF_0= \p \CN_0^{-1}=(\CN_0^{-1}\otimes \CN_0^{-1})\Delta (\CN_0)~,\nonumber \\
&&\CF_0^{-1}=\p \CN_0= \Delta (\CN_0^{-1}) (\CN_0 \otimes \CN_0)~.
\label{coboundaryrelation}
\eea
As a result, we can write the correlation function $\sigma(z,w)$ as 
\bea
\sigma(z,w)\equiv \langle\, :\!F[X]\!:\!(z) :\!G[X]\!:\!(w) \,\rangle_0
&=&\tau \circ m \left[\CF_0^{-1}
\act (F[X] \otimes G[X])
\right].
\label{F*G}
\eea
This formula is algebraically well-defined, 
where the subtraction of the divergence is already taken into account.
It contains only the divergences expected from the operator 
product of the two local operators.

Formula (\ref{F*G}) is a typical form of a twisted product triggered 
by a twist of a Hopf algebra 
(Drinfeld twist).
The main observation here is that the Wick contraction is a 
Hopf algebra action of an element $\CF_0^{-1} \in \CH\otimes \CH$.
If $\CF_0$ is a twist element, 
the product inside the path integral is given by the twisted product 
$m_{\CF_0}=m\circ \CF_0^{-1}$.
This is indeed the case as we will see below.

\subsection{Quantization as a Hopf algebra twist}
\label{sec:twist quantization}

The above discussion motivates us to regard the quantization of strings 
as a Hopf algebra twist.
In this subsection, we give a 
simple quantization procedure for defining the VEV 
on this basis, which coincides with the path integral counterpart for our example.
For a general theory of the Hopf algebra twist, see Ref.\citen{Majid} 
(see also Appendix \ref{Hopf}).

Let $\CH=U({\mathfrak X})$ be the Hopf algebra of functional vector fields and 
let $\CA$ be the algebra of classical functionals, 
which is an $\CH$-module algebra with product $m$.
Suppose that there is a twist element
(counital 2-cocycle) $\CF_0 \in \CH\otimes \CH$, that is, 
it is invertible, counital with $({\rm id} \otimes \epsilon)\CF_0=1$ 
and satisfies the 2-cocycle condition
\bea
&&(\CF_0 \otimes {\rm id})(\Delta \otimes {\rm id})\CF_0=({\rm id}\otimes\CF_0)
({\rm id} \otimes \Delta)\CF_0~.
\label{2-cocycle condition}
\eea
It is easy to show that our $\CF_0 \in \CH\otimes \CH$ (\ref{F_0}) satisfies all these conditions
(see Appendix \ref{sec:cocycle-cond}).

Given a twist element $\CF_0$, the twisted Hopf algebra $\CH_{\CF_0}$ can be
defined by the same algebra and the counit as $\CH$,
but with a twisted coproduct and antipode
\be
\Delta_{\CF_0}(h)=\CF_0 \Delta (h) \CF_0^{-1}, \quad 
S_{\CF_0}(h)=US(h)U^{-1}
\ee
for all $h \in \CH$, where $U=\mu ({\rm id}\otimes S)\CF_0$.
Correspondingly, a $\CH$-module algebra $\CA$ is twisted 
to the $\CH_{\CF_0}$-module algebra $\CA_{\CF_0}$.
It is identical to $\CA$ as a vector space 
but is accompanied by the twisted product
\be
m_{\CF_0} (F\otimes G) =m\circ \CF_0^{-1}\act (F\otimes G)~.
\label{twistedProduct}
\ee 
This twisted product is associative 
owing to the cocycle condition (\ref{2-cocycle condition}).
We also denote it as $F*_{\CF_0} G$ using a more familiar notation, 
i.e., the star product.
Note that $\CH_{\CF_0}$ is still cocommutative for our twist element $\CF_0$ (\ref{F_0}),
and thus the twisted product remains commutative.

We define the VEV for the twisted module algebra $\CA_{\CF_0}$ 
simply as the map $\tau: \CA_{\CF_0} \rightarrow \complex$ introduced in (\ref{def of tau}).
For any element $I[X] \in \CA_{\CF_0}$ the map gives
\be
\tau\left(\,I[X]\, \right).
\label{def of true VEV}
\ee
If $I[X]$ is a product of two elements in $\CA_{\CF_0}$,
using the above notation, their correlation function $\sigma(z,w)$ follows from 
(\ref{def of true VEV}) as
\be
\sigma(z,w)=\tau(F[X(z)]*_{\CF_0}G[X(w)]),
\ee
which coincides with the path integral version of 
$\sigma(z,w)$ in (\ref{F*G}) for $\CF_0$ in (\ref{F_0}).
Because the cocycle condition guarantees the associativity of the twisted product,
the correlation function of $n$ local functionals is similarly 
\be
\sigma(z_1,...,z_n)=\tau(F_1[X(z_1)]*_{\CF_0}F_2[X(z_2)]\cdots *_{\CF_0}F_n[X(z_n)]),
\ee 
which again coincides with the path integral.
Therefore, for the twist element $\CF_0$ in (\ref{F_0}), 
this process of twisting is identical with
the path integral.
We emphasize that the process does not depend on the action $S_0$ but only on the twist 
element $\CF_0$.
Moreover, the twist element in our example $\CF_0$ in (\ref{F_0}) 
is only accompanied by the Hopf subalgebra of the worldsheet variations. 
Indeed, $\CF_0 \in U({\mathfrak C})\otimes U({\mathfrak C})$ 
and it is determined by the free propagator $G_0^{\mu\nu}(z,w)$.
It is then easy to generalize our twist element to more general twist elements
of the same form as (\ref{F_0}) but with different Green functions $G_0^{\mu\nu}(z,w)$.
They correspond to different worldsheet theories with quadratic actions $S_0$.
Our proposal is that 
given a Hopf algebra $\CH$ and a module
algebra $\CA$ defined in terms of classical 
functionals as in the previous section, 
then for {\sl any} twist element,  
the resulting twisted Hopf and module algebras
give a quantization on the worldsheet.
A different choice of the twist element gives a different 
quantization scheme.
We will come back to this point in the next section 
from the viewpoint of the space-time symmetry.

It is instructive at this 
stage to compare the twisted product $*_{\CF_0}$
in this paper and the star product in deformation quantization \cite{BFFLS} 
in quantum mechanics,
because they share the same property.\footnote{
In the case that the phase space is a Poisson-Lie group, this deformation is
equivalent to the Hopf algebra twist of the universal enveloping algebra of the 
dual Lie algebra $U_{\hbar}({\mathfrak g})$ \cite{Drinfeld}.}  
Both theories can be described by 
classical variables even after the quantization.
The latter is generalized to field theories \cite{Dito} 
and also to string theory \cite{GarciaCompean}.
We will discuss this point in a separate paper and 
do not develop it further here, but
a few remarks about this issue are in order.

In deformation quantization, a classical Poisson algebra of observables on the phase space
is deformed by replacing its commutative product with a star product.
It is accompanied by a (formal) deformation parameter $\hbar$ such that 
in the limit $\hbar \rightarrow 0$ the undeformed algebra is recovered. 
A basic example is the phase space $\real^{2n}$ equipped with 
a symplectic structure $\omega$,
where the algebra $C^{\infty}(\real^{2n})$ of complex functions 
is extended to $C^{\infty}(\real^{2n})[[\hbar]]$, a formal 
power series in $\hbar$,
and the product is twisted by
$
e^{- \frac{i}{2} \hbar \sum \omega^{ij}\p_i \otimes \p_j}
$,
where $\omega^{ij} $ is the inverse of a symplectic matrix.
This algebra with the star product corresponds 
to the operator formulation in quantum mechanics 
in the Schr\"odinger picture, 
where the information on the time evolution is contained in the wave function.
This correspondence is essentially the same for deformation quantization in field theories.

Comparing this with the twist $\CF_0$ (\ref{F_0}), formally,
the propagator $G_0^{\mu\nu}(z,w)$ plays the role of $\hbar \omega^{ij}$.
However, we should keep in mind the following differences:
to be explicit, we take the worldsheet $\Sigma=\complex$ for simplicity.
In this case, the propagator $G_0^{\mu\nu}(z,w)$ is given by \eqref{tree propagator}.
First, a twist can also be accompanied by a deformation 
parameter.
In our case, it is $\alpha'$, because the loop expansion parameter 
in front of the action is $\alpha'$ (we fix $\hbar=1$).
The Hopf algebra $\CH$ and the
module algebra $\CA$ are considered to be already 
extended to include $\alpha'$ by dimensional reasoning.
Therefore, the generic elements of 
$\CA_{\CF_0}$ can contain $\alpha'$ and the 
twisted product gives a power series in $\alpha'$ relative to these.
Second, our twisted product depends on the dynamical evolution (it is free)
on the worldsheet, which is more alike to the Heisenberg picture 
in quantum mechanics.
This explains the missing factor of $\ln|z-w|$ in the deformation 
quantization, 
and therefore the remaining $\eta^{\mu\nu}$ correspond to $\omega^{ij}$. 
From the viewpoint of the target space-time, it would be 
better to regard a deformation 
parameter as $\alpha'\times$(worldsheet distributions).
For a given target space $\real^d$ and a (choice of) background metric $\eta_{\mu\nu}$,
we have a twisted product defined by $\CF_0$.
Third, there is an important difference between our twist and that
 of the deformation quantization.
In the latter the factor in the exponential is antisymmetric under the interchange of 
the partial derivatives, while $\CF_0$ is symmetric in this sense.
This implies that the twist $\CF_0$ is formally trivial, in contrast to the deformation quantization.

\subsection{Normal ordering}
\label{sec:normal ordering}

Here we clarify the meaning of the formula for the VEV in 
the path integral   
\eqref{VEVpathintegral} and in the r.h.s. of \eqref{F*G}.
Recall that our twist $\CF_0$ can be written in terms of 
$\CN_0$ as (\ref{coboundaryrelation}).
This is also the case for any twist element 
$\CF_0 \in U({\mathfrak C})\otimes U({\mathfrak C})$
of the form (\ref{F_0}).
From the viewpoint of the Hopf algebra cohomology,
this means that the twist element $\CF_0$ is a coboundary 
and thus it is trivial.
See Appendix \ref{trivial}. (There, we set $\CH_\chi =\CH, \chi=1\otimes 1,\CH_\psi=\CH_{\CF_0}, 
\gamma={\CN_0}^{-1} \in \CH$).
Then, there is an isomorphism between the Hopf algebras
 $\hat{\CH}$ and $\CH_{\CF_0}$ 
(module algebras $\hat{\CA}$ and $\CA_{\CF_0}$ respectively) summarized 
as
\bea
\begin{array}{ccccc}
\CH & \xrightarrow{\mbox{twist}} & \CH_{\CF_0} & \xrightarrow{\sim} & 
\hat{\CH} \\
\triangledown & & \triangledown & & \triangledown  \\
\CA & \xrightarrow{\mbox{twist}} & \CA_{\CF_0} & \xrightarrow{\sim} 
& \hat{\CA}  
\end{array}
\label{diagram}
\eea
In the diagram, the left row is a classical pair $(\CH,\CA)$, 
and the middle and right rows are their quantum counterparts.
Here the map $\CH_{\CF_0} \xrightarrow{\sim} \hat{\CH}$ is given by the 
inner automorphism $h \mapsto \CN_0 h {\CN_0}^{-1} \equiv \tilde{h}$, and 
the map $\CA_{\CF_0} \xrightarrow{\sim} \hat{\CA}$ is given by 
$F \mapsto \CN_0 \act F \equiv :\!F\!:$.
We call $\CH_{\CF_0}$ ($\CA_{\CF_0}$) 
the twisted Hopf algebra (module algebra) while
$\hat{\CH}$ ($\hat{\CA}$) is called the normal 
ordered Hopf algebra (module algebra).
The reason why we distinguish between the classical $(\CH,\CA)$ and the 
normal ordered 
 $(\hat{\CH},\hat{\CA})$ pairs 
(they are formally the same) is explained below.

To understand the physical meaning of this diagram,
let us focus on the module algebras
(we will discuss the Hopf algebra action in the next section).
Since a functional $F \in \CA_{\CF_0}$ is mapped to $\CN_0 \act F \equiv :\!F\!:$, 
the elements in $\hat{\CA}$ are normal ordered functionals.
The VEV (\ref{def of true VEV}) for $\CA_{\CF_0}$ implies that we should 
identify (\ref{VEVpathintegral}) as the definition of the VEV for $\hat{\CA}$, i.e., the map
$\tau \circ \CN_0^{-1}: \hat{\CA}\rightarrow \complex$.
The product in $\CA_{\CF_0}$ is mapped to that in $\hat{\CA}$:
\be
 \CN_0 \act m \circ \CF_0^{-1} \act (F\otimes G) 
= m \circ (\CN_0 \otimes \CN_0)\act (F\otimes G)~,
\label{product iso}
\ee
which is a direct consequence of the coboundary relation (\ref{coboundaryrelation}).
An equivalent but more familiar expression 
$:\! (F *_{\CF_0} G) \!: = :\!F\!:\, :\!G\!:$
is simply (\ref{Pol-identity}), the time ordered product of the vertex operators, which is again equivalent to (\ref{F*G}) in the path 
integral average
\bea
\langle\, :\!F[X]\!:\!(z) :\!G[X]\!:\!(w) \,\rangle_0
&=& \langle \, :\! F[X](z) *_{\CF_0} G[X](w) \!: \, \rangle_0~.
\eea
From these considerations, 
all the quantities and operations 
in the path integral average, such as in the l.h.s. of (\ref{F*G}) should be 
understood as the objects in 
the normal ordered Hopf algebra and module algebra.
The isomorphism implies that, formally, the quantization is performed 
either 
by a twist $\CF_0$ or by changing the element determined by $\CN_0$ 
in the path integral.
The latter corresponds to the operator formulation.

However, there are some differences between the twist 
quantization and the path integral in the following sense.
Note that the twist from $\CA$ to $\CA_{\CF_0}$ 
changes the product but it does not change the elements.
Therefore,  a classical functional $F$ does not suffer from a 
quantum correction ($\alpha'$-correction) under the twist.
On the other hand, the map $\CA \rightarrow \hat{\CA} : F \mapsto :\!F\!:$ 
changes the elements while it does not change the operations.
Because $\CN_0$ contains $\alpha'$, the normal ordered functional 
$:\!F\!:$ is necessarily a power series 
in $\alpha'$ (relative to $F$) and each term in the series  
is always divergent because of the propagator at the coincident point.
Therefore, $:\!F\!:$ should be distinguished from the classical functional $F$. 
Nevertheless,
this {\it does not} mean that the normal ordered module algebra $\hat{\CA}$ is ill-defined,
rather one should think of it as an artifact of 
the description, which is based on 
the classical functional. 
In fact, in the path integral, the normal ordered functionals 
give finite results 
but the classical functional is divergent.
There is a similar argument in the 
deformation quantization approach to field theories:
only the normal ordered operator corresponding to this divergent functional 
is well-defined within the canonical quantization \cite{Dito},
while a Weyl ordered operator corresponding to a classical functional 
has a divergence due to the infinite zero point energy.
In this sense, if we adopt the description based on 
the normal ordering of operators,
$\hat{\CA}$ is the natural object and is well-defined in the path integral average.

However, if we consider a different 
choice of the background in string theory, 
there is a significant difference between twisted and normal ordered descriptions.
The latter is highly background-dependent, because both, the element 
$:\!F\!:\in \hat{\CA}$
and the VEV $\tau \circ \CN_0^{-1}$ contain $\CN_0$.
As seen in the example of the propagator in \eqref{tree propagator}, $\CN_0$ depends on
the background metric $\eta_{\mu\nu}$. 
This corresponds in the operator formulation to the property that 
a mode expansion of the string variable $X^\mu(z)$ and 
the oscillator vacuum are background-dependent.
Therefore, the description of the quantization 
that makes $\hat{\CA}$ well-defined is 
only applicable to that background and we need another mode expansion for another background.
On the other hand, elements in twisted Hopf and module algebras 
are not altered,
thus they have a background-independent meaning.
All the effects are controlled by only the single twist element $\CF_0$; 
thus, the background dependence is clear.
In this respect, we can claim that the quantization as a Hopf algebra twist is 
a more general concept than the ordinary treatments.
One of the advantages of this viewpoint will become clearer 
when we consider the space-time 
symmetry in the next section.

We finish this subsection with a remark: 
A Hopf algebra structure underlying the Wick 
contraction and the normal ordered product 
has been already considered in the literature. \cite{BrouderOeckl}\cite{BFO}
In their approach, the algebra with a normal ordered product was an untwisted 
Hopf algebra (symmetric algebra) and, by twisting with the propagator 
(Laplace pairing) the twisted module algebra became an algebra 
with a time ordered product.
One difference between Ref.\citen{BFO} and our treatment is 
that the approach in the former is based on the mode expansion.
The approach in Ref.\citen{BFO} may be related to ours 
but we do not discuss the details here in this paper.

\section{Space-time symmetry}
\label{sec:symmetry}

In the previous section, 
we formulated the quantization as a twist of a Hopf algebra. 
The VEV of a product of local vertex operators was 
formulated as a twisted product $*_{\CF_0}$ of functionals 
in the module algebra and the map $\tau$. 
The twist of the module algebra was a consequence of the 
twist of the Hopf algebra acting on the classical local functionals.
Here we focus on the twisted Hopf algebra itself, in particular 
on its relation with the space-time symmetry.
After discussing the general structure of the twisted Hopf algebra,
we see how the diffeomorphism is realized in a fixed background.
We also give identities among correlation functions, such as the Ward identity.

\subsection{Twisted Hopf algebra and its action}
\label{sec:twisted Hopf algebra}

Here, we continue to describe the process of twisting discussed 
in \S \ref{sec:twist quantization}.
We start with describing the effect of  
the twist $\CH \rightarrow \CH_{\CF_0}$ 
acting on the module algebra, then 
we discuss the (formal) isomorphism 
$\CH_{\CF_0}\simeq \hat{\CH}$ in (\ref{diagram}).

An action of an element $h\in \CH$ on a classical functional 
$h\act I[X] \in \CA$ represents a variation 
under a classical transformation 
(diffeomorphism or worldsheet variation).
The twist element $\CF_0$ causes a twisting of the Hopf algebra $\CH \rightarrow \CH_{\CF_0}$,
and the consistency of the action (i.e., covariance) 
requires that the twisted functional algebra $\CA_{\CF_0}$ 
is again an $\CH_{\CF_0}$-module algebra.
Since each element in $\CH_{\CF_0}$ as well as in $\CA_{\CF_0}$ 
is the same as the corresponding classical element, 
the variation of the local functional has 
the same representation $h \act F[X]$ as the classical transformation.
However, since the coproduct is deformed into 
$\Delta_{\CF_0}(h)=\CF_0 \Delta(h) \CF_0^{-1}$, the action 
is not the same as the classical transformation when $I[X]$ is a product 
of several local functionals.
The covariance of the twisted action on the twisted product (\ref{twistedProduct}) of two functionals in $\CA_{\CF_0}$ is guaranteed
by the covariance of the original module algebra (\ref{covariance}) as
\bea
 h\act m_{\CF_0} (F\otimes G)
&=& h \act m\circ \CF_0^{-1}\act (F\otimes G) \nonumber\\
&=& m\circ \Delta (h) \CF_0^{-1}\act (F\otimes G) \nonumber\\
&=& m \circ \CF_0^{-1} \Delta_{\CF_0} (h) \act (F\otimes G) \nonumber\\
&=& m_{\CF_0} \Delta_{\CF_0} (h) \act (F\otimes G)~.
\label{twisted covariance}
\eea
In this way the Hopf algebra and the module algebra are twisted in a consistent manner.

From the viewpoint of quantization, the twisted module algebra $\CA_{\CF_0}$
together with the map $\tau: \CA_{\CF_0} \rightarrow \complex$ defines a VEV in a 
quantization of the string worldsheet theory.
Then, the twisted Hopf algebra $\CH_{\CF_0}$ should be regarded as 
a set of quantum symmetry 
transformations, which is consistent with the quantized (twisted) product.
In other words, classical space-time symmetries should 
also be twisted under the 
twist quantization.
The corresponding variation inside the VEV $\tau (I[X])$ 
is given by
\be
\tau \left(h \act I[X]\right)
\label{Hopf action}
\ee
and this appears in the various relations involving the symmetry transformation.

Next recall the (formal) isomorphism 
$\CH_{\CF_0}\simeq \hat{\CH}$ in (\ref{diagram})
(see also Appendix \ref{trivial}).
Under the isomorphism map, $F\xrightarrow{\sim} :\!\!F\!\!:\,=\CN_0\act F$ of module algebras, 
the action of $h \in \CH_{\CF_0}$ on $\CA_{\CF_0}$ is 
mapped to the action of $\tilde{h}= \CN_0 h \CN_0^{-1} \in \hat{\CH}$ on $\hat{\CA}$ as
\be
h\act F ~\xrightarrow{\sim}~ 
\CN_0 \act (h \act F) =\,
\tilde{h}~\act :\!\!F\!:.
\label{action on :F:}
\ee
Correspondingly, the action on the product is
$h\act (F*_{\CF_0} G) \xrightarrow{\sim}\tilde{h} \act (:\!F\!::\!G\!:)$.
The covariance of the $\hat{\CH}$-action on $\hat{\CA}$ 
can be proven by applying $\CN_0$ to both sides of (\ref{twisted covariance}):
\bea
\tilde{h} \act (:\!F\!::\!G\!:)
&=& \CN_0 h \act (F*_{\CF_0} G) \nonumber\\
&=& m \circ \Delta (\CN_0 h) \CF_0^{-1} \act(F\otimes G) \nonumber\\
&=& m \circ \Delta (\tilde{h}) (\CN_0\otimes \CN_0) \act (F\otimes G)~~.
\eea
As argued in \S \ref{sec:quantization}, some elements in the normal ordered algebra 
contain the formal divergent series in the functional language, 
and thus the above equation has only a meaning under the path integral. 
For example,  for a single local insertion, (\ref{Hopf action}) leads to
\bea
\langle\, \tilde{h} ~\act :\!F[X]\!:(z) \,\rangle_0
&=&
\tau \left( h \act F[X(z)] \right),
\label{Hopf action for F}
\eea
and for a product of local functionals
\bea
\langle\, \tilde{h} \act (:\!F[X]\!:\!(z) :\!G[X]\!:\!(w) ) \,\rangle_0
&=&
\tau \left( h \act m( \CF_0^{-1} \act (F[X] \otimes G[X])) \right).
\label{Hopf action for F*G}
\eea
In this way it is always possible to convert the action of the 
twisted Hopf algebra into that of the normal ordered Hopf algebra.
However, we will see below that the structure of the diffeomorphism 
is far simpler written in terms of the twisted Hopf algebra than in terms of 
the normal ordered Hopf algebra.
Related to this, another
way to give a well-defined meaning to $:\!F\!:$ is to replace it 
by the normal-ordered operator. In this case, the action $\tilde{h}\,\act $ should also 
be replaced with an operation involving operators and it becomes 
strongly background-dependent
owing to $\CN_0$ being included in the definition of $\tilde{h}$.

As we have seen in \S \ref{sec:quantization}, 
the quantization itself is performed within a worldsheet twist,
namely, $\CF_0$ in (\ref{F_0}) is a twist element of a Hopf subalgebra 
$U({\mathfrak C})$.
However, this twist affects the whole Hopf algebra
 $\CH=U({\mathfrak X})$.
This can be understood as follows:
Any classical vector field $\xi$ is originally primitive, 
$\Delta (\xi)=\xi\otimes 1+1\otimes \xi$, that is, it obeys the Leibniz rule.
After twisting, $\xi$ in $\CH_{\CF_0}$ acts 
on a single functional $F$ in the same way, 
$\xi \act F$, as in the classical case, but it is not, in general, 
primitive now, 
since the coproduct is twisted 
$\Delta_{\CF_0}(\xi)=\CF_0 \Delta (\xi)\CF_0^{-1}$.
This occurs when $\CF_0$ does not commute with 
$\Delta (\xi)$.\footnote{
By the isomorphism, 
the same property holds for the normal ordered Hopf algebra $\hat{\CH}$.
An element $\tilde{\xi}=\CN_0 \xi \CN_0^{-1}$ is not primitive 
unless $[N_0,\xi]=0$ owing to the factor $\Delta (\CN_0)$.
}
Therefore, as a consequence of the twist quantization,
the diffeomorphism of space-time, in general, cannot be separately 
considered but it should be twisted as well.

In this respect, the universal enveloping algebra $U(\CP) \subset \CH$ 
for the Poincar\'e Lie algebra $\CP$ is a special case, 
since even after the twisting, it is identical with the original $U(\CP)$.
This can be seen by proving that the twist does not alter the 
coproduct $\Delta_{\CF_0} (u)=\Delta (u)$ 
as well as the antipode $S_{\CF_0} (u)=S(u)$
for ${}^\forall u \in U(\CP)$.
See Appendix \ref{sec:Poincare} for the proof. 
Therefore, $U(\CP)$ is also a Hopf subalgebra 
of $H_{\CF_0}$.

With our choice of the twist element $\CF_0$ (\ref{F_0}), 
we argued that the twist quantization coincides with the ordinary quantization,
in which the Poincar\'e covariance is assumed to be at the quantum level.
This suggests that, in general, the twist-invariant Hopf subalgebra corresponds to the 
unbroken symmetry, while the full diffeomorphism should be twisted 
under the quantization of the chosen twist element.
Below we elaborate on the physical meaning of the twisted Hopf algebra
from the viewpoint of background-(in)dependence.
In particular, in the following subsections, 
we discuss the meaning of the twisted diffeomorphism in this context
and the characterization of the broken/unbroken symmetries
together with the relation with the various identities among the 
correlation functions.

\paragraph{\it Identities in path integrals}

Before we discuss the various identities related to the symmetries
in the twisted Hopf algebra, 
we recall the ordinary path integral relations 
for the symmetry transformations.
In the path integral, the identities are obtained 
using the fact that any change of variables gives the same result.
In particular, under the constant shift 
$X^\mu (z) \mapsto X^\mu (z)+ \varepsilon^\mu$, 
where $\varepsilon^\mu$ is a constant, 
it gives the identity
\be
\label{path int total derivative}
0= \int \CD X {\delta \over \delta X^\rho (z)} \left( e^{-S_0}\CO \right)~.
\ee
More generally, consider an arbitrary infinitesimal change of variables 
$X^{'\mu}(z)=X^\mu (z)+\xi^\mu (z)$.
Then,
\bea
0&=& \int \CD X' e^{-S_0[X']}\CO[X'] - \int \CD X e^{-S_0[X]}\CO[X] \nonumber \\
&=& \int \CD X e^{-S_0[X]} \left\{
J \CO -(\xi \act S_0) \CO + \xi \act \CO
\right\} ,
\label{general variation}
\eea
where the first term is the Jacobian obtained from the variation 
of the measure $\CD X' := \CD X(1+J)$, 
and the second term is the 
variation of the action, $S_0[X']=S_0[X]+\xi \act S_0[X]$.\footnote{
It is related to the variation 
$\delta_\xi S_0=-\xi\act S_0$. 
See also \S \ref{sec:classical}.
}.
Here we have used Hopf algebra notation 
for the action of the functional derivative.

We can derive various identities from \eqref{general variation} as follows:
\begin{enumerate}
\item[(i)] If $\xi$ generates a worldsheet variation independent of $X$ 
(i.e., $\xi \in {\mathfrak C}$), 
then the measure is manifestly
invariant, $\CD X' = \CD X$, and \eqref{general variation} reduces to
\be
0= \int \CD X e^{-S_0}\left\{ 
-(\xi \act S_0) \CO + \xi \act \CO
\right\} ,
\label{SDeq}
\ee
which is used to derive the Schwinger-Dyson equation.
\item[(ii)] The case $\xi$ generates a space-time symmetry: 
If $\xi=\xi[X(z)]$ is a target space vector field, 
but the measure and the action are invariant under $\xi$, then 
only the third term remains, and we obtain the Ward identity
\be
0= \int \CD X e^{-S_0}\left\{ 
  \xi \act \CO
\right\} ,
\ee
where the transformation acts only on the insertions $\CO$.
\item[(iii)] Eq. \eqref{general variation} is also used to derive
Noether's theorem in the path integral language 
(see, for example, Ref.\citen{Polchinski}).
Under the same assumption as stated in (ii), but extending the variation 
to $X'=X+ \rho(z) \xi$ by an arbitrary 
distribution $\rho(z)$ on the worldsheet, 
the measure is still invariant. However
 the variation of the action is written as 
$\int J^a \p_a \rho (z)$, where $J^a$ is the Noether current,
leading to the identity 
\be
0= \int \CD X e^{-S_0} \rho(z) \left\{ 
 -\left(\int dS_a J^a  \right) \CO + \xi \act \CO
\right\}~.
\ee
Note that the insertion of the Noether current (charge) is 
written as the classical variation of the operator insertion.
Thus, the above equation corresponds to the operator identity of the symmetry transformation 
at the quantum level,
$[Q, {\cal O}]=\delta {\cal O}$,
where in the l.h.s., $Q$ is the generator of the transformation 
and the r.h.s. is the classical 
variation of the operator ${\cal O}$.
\item[(iv)] If the vector field does not preserve the action $S_0$, 
it is not a classical symmetry but (\ref{general variation}) still 
represents a 
broken Ward identity that incorporates the change of the action $S_0$ (and the measure).
\end{enumerate}

In the following we derive the same 
type of identities in the Hopf algebra language, 
where the path integral VEV is replaced by the algebraic 
operation $\tau (I[X])$. 
Before we start the derivations, we point out
the following:
The variation in the integrand in (\ref{general variation}) is the 
sum of that of the action $S_0$ and of each insertion.
In particular, the action $\xi \act {\cal O}$ 
on the multiple insertions satisfies the Leibniz rule.
It states that formally the vector field $\xi$ should be an 
element of the classical Hopf algebra
$\CH$.\footnote{Although the VEV itself belongs to the quantum theory, the variation is classical, and the transformed classical functional is 
integrated giving a new VEV.}
However, at the same time, it is
implicitly assumed that each insertion is understood to be normal 
ordered one in the path integral method.
Therefore, we should be careful when this normal ordering is applied.  
In other words, we need to understand the change of variables in the 
Hopf algebra language to characterize the path integral 
identities correctly.

\subsection{Twisted Hopf algebra as a symmetry}
\label{sec:change of variables}

We clarify the relation between the Hopf algebra action of the twisted 
$\CH_{\CF_0}$ (normal ordered $\hat{\CH}$), the Hopf algebra
described in the previous subsection 
and the change of variables representing a symmetry
transformation.
Although they are identical classically, the relation
is not trivial after the quantization. 

Let $\xi \in {\mathfrak X}$ be a vector field and let 
$u=e^{\xi} \in \CH=U({\mathfrak X})$ be 
an element of the classical Hopf algebra.
It is a group like element, $\Delta (u)=u\otimes u$
and it acts on both $\CA$ and $\CH$.  
Its action on the variable $X$ defines $X'^\mu=u \act X^\mu$ as a 
new variable.
Then its action on any functional, $u \act I[X]=I[X']$, is considered to be 
the transformation law caused by the change of variables\footnote{
The variation $\delta_\xi I[X]$ 
is still defined infinitesimally $\xi\sim 0$,
and is written by the $u$-action as
$\delta_\xi I[X]=-\xi \act I[X] = u^{-1} \act I[X]|_{{\cal O}(\xi)}=-I[X']|_{{\cal O}(\xi)}$.
Here $\xi\sim 0$ and ${\cal O}(\xi)$ are abbreviations, 
that are more rigorously defined by the $t\rightarrow 0$ limit  
of the action of the flow $u(t)=\exp{(t\xi)}$ generated by $\xi$.
Note also that the relation is valid bacause $I[X]$ is a scalar functional.
}.
Of course, $I[X']$ is also an element of the classical 
functionals $\CA$.
Since $u$ is group like, the transformation law for the product of functionals
is the product of each functional.
In particular, the classical diffeomorphism is given by $u=e^{\xi}$ 
with $\xi[X]$ being 
the  pull-back of a space-time vector field.

Note that the adjoint action of $\xi$ on $h \in \CH$ is defined by the Lie bracket $[\xi, h]$. 
Then the action of $u$ on the functional derivative gives  
\bea
{\delta \over \delta X^{'\mu} (z)} 
&:=& u\act {\delta \over \delta X^{\mu} (z)} \nonumber \\
&=& \sum_{n=0}^\infty {1\over n!} \xi^n \act {\delta \over \delta X^{\mu} (z)} \nonumber \\
&=& e^{\xi} {\delta \over \delta X^{\mu} (z)} e^{-\xi}  \nonumber \\
&=& u {\delta \over \delta X^{\mu} (z)} u^{-1},
\label{generalized chain rule}
\eea
which is a finite version of the chain rule under the change of variables.
The functional derivative is transformed such that the relation 
\be
{\delta \over \delta X^{'\mu} (z)} \act X^{'\nu}(w) 
= u {\delta \over \delta X^{\mu} (z)} u^{-1} u\act X^\nu(w) 
= \delta^\nu_\mu \delta^{(2)}(z-w) 
\ee
is maintained after the change of variables.
${\delta \over \delta X^{'\mu}}$ can be used to construct ${\mathfrak X}$
just as the original functional derivative do.
The transformed functional derivatives are primitive since 
\be
\Delta \left({\delta \over \delta X^{'\mu}}\right)=
(u\otimes u)\Delta \left({\delta \over \delta X^{\mu}}\right)(u^{-1}\otimes u^{-1}) 
= {\delta \over \delta X^{'\mu}}\otimes 1+1\otimes {\delta \over \delta X^{'\mu}}.
\label{primitive}
\ee
Their commutator vanishes, 
\bea
\left[{\delta \over \delta X^{'\mu} (z)}\,, {\delta \over \delta X^{'\nu} (w)} \right]
&=& u \left[{\delta \over \delta X^{\mu} (z)}\,, {\delta \over \delta X^{\nu} (w)} \right] u^{-1}
= 0.
\label{commutes}
\eea
It is straightforward to show that 
for the element $h \in U({\mathfrak C})$ we 
have $h'=u\act h = u h u^{-1}$, which is equivalent to replacing
 all the functional 
derivatives in $h$ with ${\delta \over \delta X^{'\mu}}$.
For example, $\CN'_0=u\CN_0 u^{-1}$.
For an arbitrary element $h \in \CH$, $u$ also acts on the coefficient function.

Let $\CF_0$ be a fixed twist element in $U({\mathfrak C})\otimes U({\mathfrak C})$.
The action of $u$ on $\CF_0$ is written as
\bea
\CF'_0 &=& u \act \CF_0 \nonumber \\
&=& \Delta (u) \CF_0 \Delta (u^{-1}) \nonumber\\
&=& (u\otimes u)\CF_0 \Delta (u^{-1}).
\label{group coboundary}
\eea
This means that $u$ is a coboundary (this is always the case for a 
group like element)
in a cohomological sense 
and  $\CF'_0$ is a new twist element equivalent to $\CF_0$ (see Appendix \ref{trivial}). 
It is also easy to show directly that $\CF'_0$ satisfies the cocycle condition by using 
the expression 
\be
{\CF'}_0
= \exp\left\{-\int \! d^2z \!\! \int \! d^2w \,
G_0^{\mu\nu}(z,w){\delta\over\delta X^{'\mu}(z)}
\otimes {\delta\over\delta X^{'\nu}(w)}\right\},
\label{F'_0}
\ee
since the proof depends only 
on the properties of the functional derivatives 
${\delta \over \delta X^{'\mu}}$ in (\ref{primitive}) and (\ref{commutes}) 
(see Appendix \ref{sec:cocycle-cond}).

Now consider the action of $u$ on the twisted module algebra $\CA_{\CF_0}$. 
For any functional $I[X] \in \CA_{\CF_0}$, its transformation law is 
the same as that of $\CA$, $I[X']=u\act I[X]$.
This is also true for the star product of two local functionals 
$I[X]=F[X]*_{\CF_0}G[X]$ 
when it is considered as a functional of $X$ after the star 
product is performed.
Because of (\ref{twisted covariance}), this action of $u$ is nothing
but the twisted Hopf algebra action as
\bea
u\act (F[X] *_{\CF_0}G[X]) &=& m \circ \Delta (u) \CF_0^{-1} \act (F\otimes G) \nonumber\\
&=& m \circ \CF_0^{-1} \Delta_{\CF_0}(u) \act (F\otimes G).
\label{twist viewpoint}
\eea
However, the same action can also be 
written using (\ref{group coboundary}) as 
\bea
u\act (F[X] *_{\CF_0}G[X]) &=& m \circ \Delta (u) \CF_0^{-1} \act (F\otimes G) \nonumber\\
&=& m \circ {\CF'}_0^{-1} (u\otimes u) \act (F\otimes G) \nonumber \\
&=& F[X'] *_{\CF'_0}G[X'].
\label{another viewpoint}
\eea
The r.h.s. is equivalent to the replacement of each $X$ with $X'$ as well as each
${\delta \over \delta X^{\mu}}$ with ${\delta \over \delta X^{'\mu}}$ before the 
star product is performed.
This gives a good understanding of the twisted diffeomorphism.\footnote{
There is essentially the same argument in the context of noncommutative gravity in Ref.\citen{AG}.
In fact, (\ref{another viewpoint}) can be rewritten in terms of the 
variation by using the relation mentioned in the previous footnote.
Then we obtain 
$\delta_\xi (F*_{\CF_0} G)[X]=(\delta_\xi F*_{\CF_0} G)[X] + (F*_{\CF_0} \delta_\xi G)[X] 
+(F *_{\delta_\xi \CF_0} G)[X]$, which is the formula in Ref.\citen{AG}.
Here, the product in the third term is defined by inserting 
$\delta_\xi \CF_0^{-1} = -{\CF'}_0^{-1} |_{{\cal O}(\xi)} 
= -[\Delta(\xi),\CF_0^{-1}]$.}
Comparing these two expressions we see that the action of $u$ as the 
twisted Hopf algebra (\ref{twist viewpoint}) 
is simply the classical diffeomorphism in which the twist element 
has also been transformed (\ref{another viewpoint}).
Here the change of the twist element itself is converted to the change of functionals through 
the twisted coproduct $\Delta_{\CF_0}$ while keeping the 
twist element invariant.

Moreover, (\ref{another viewpoint}) is seen as a product in the new twisted module algebra 
$\CA_{\CF'_0}$ twisted by $\CF'_0$.
From (\ref{group coboundary}), the new twisted algebra 
is isomorphic to the 
original one. 
Denoting this isomorphism as $\rho : \CA_{\CF_0}\rightarrow \CA_{\CF'_0}$, 
given by $\rho (F)= u \act F$, then
(\ref{another viewpoint}) implies that $\rho (F*_{\CF_0}G)= \rho (F)*_{\CF'_0}\rho (G)$.
This gives a new viewpoint for the twist, 
that is, a change of background under $\rho$ relates 
two twists $\CA\rightarrow\CA_{\CF_0}$ and $\CA\rightarrow\CA_{\CF'_0}$ with each other.
It is equivalent to regard the new twist element (\ref{F'_0}) as
\bea
&&\CF'_0:=\exp\left\{-\int \! d^2z \!\! \int \! d^2w \,
{G'}_0^{\mu\nu}(z,w){\delta\over\delta X^\mu(z)}
\otimes {\delta\over\delta X^\nu(w)}\right\}~~,\nonumber\\
&&{G'}_0^{\mu\nu}(z,w)= G_0^{\mu\nu}(z,w) 
-\p_\rho \xi^\mu (z) G_0^{\rho \nu}(z,w)- G_0^{\mu\rho }(z,w)\p_\rho \xi^\nu (w),
\label{F'_0 rewrite}
\eea
where the change of the propagator coincides with the transformation 
of our fixed background metric 
$\eta^{\mu\nu}$ under the diffeomorphism $u^{-1}=e^{-\xi[X]}$. 
From this viewpoint, an element $u=e^{\xi[X]}$ 
of a space-time diffeomorphism 
that keeps the twist element invariant, $\CF'_0=u\act \CF_0=\CF_0$,
is a symmetry (isometry) in the ordinary sense.

From the quantization viewpoint, we should
fix a twist element $\CF_0$ to quantize the theory.
Accompanied by this element, the classical Hopf algebra becomes a
 twisted Hopf algebra $\CH_{\CF_0}$ 
acting as  a quantized transformation.
Since a particular metric is chosen by fixing the twist, a full 
diffeomorphism is not manifest from the symmetry viewpoint. 
However, the above argument shows that the twisted diffeomorphism is 
a remnant of the classical diffeomorphism.
This is the essence of our proposal, that the twist governs the quantization 
and the space-time symmetry in a consistent way.
A fixed twist element determines the quantization scheme as well as 
the background metric,  
but the diffeomorphism is retained as a twisted Hopf algebra.
This is a good starting point for discussing the general covariance 
by extending the above argument.

\paragraph{\it Vacuum expectation value} 

We now consider the effect on the 
VEV (\ref{def of true VEV}) under the change of variables $X\rightarrow X'$, 
as in the path integral argument (\ref{general variation}).
For a functional in $\CA_{\CF_0}$, it was shown above that $I[X']=u\act I[X]$.
Next, $\tau$ is replaced with $\tau'$, which sets $X'=0$ in the functional $I[X']$.
One can show that $\tau'$ is written as an operation on $X$ as 
\be
\tau' = \tau \circ u^{-1} \act.
\label{tau'}
\ee
Combining these contributions, we have the identity 
\be
\tau' (\,I[X']\,) 
=\tau \left(u^{-1} \act (u\act I[X])\right)
=\tau (\,I[X]\,),
\label{trivial identity}
\ee
which implies that the change of variables keeps the VEV invariant.

Equation (\ref{trivial identity}) as well as (\ref{VEV=VEV}) below 
are desirable properties of the VEV corresponding to the path integral,
and they can be used to obtain various identities.
However, it appears to be difficult to derive (\ref{general variation}) directly from them,
even if we restrict ourselves to the infinitesimal change $\xi\sim 0$,
because the present formulation does not use the action 
$S_0$ transparently.

\paragraph{\it Effects on $\hat{\CA}$}

In a similar manner we can estimate in a similar manner the effect of the change of variables
on the normal ordered module algebra $\hat{\CA}$.
However, the situation is somewhat 
different from that of the twisted module algebra.
This is because the former is highly background-dependent.

The action of an element $u=e^{\xi}$ of the classical Hopf algebra $\CH$ 
on the normal ordering element is given by $\CN'_0 = u\act \CN_0 = u\CN_0 u^{-1}$.
Then, a single local vertex operator $:\!F[X]\!:=\CN_0 \act F[X] \in \hat{\CA}$ is transformed 
under the classical diffeomorphism in the same manner as (\ref{another viewpoint}),
\be
\CN'_0 \act F[X'] 
= u \CN_0 \act F[X] 
= u\,\act :\!F[X]\!:~,
\label{change of normal ordered}
\ee
which is simply a classical action of $u$ when $:\!F[X]\!:$ is considered as a functional 
of $X$ after the normal ordering is performed.
Note that this is not the $\hat{\CH}$-action 
$\tilde{u}\,\act :\!F[X]\!: = :\!u\,\act F[X]\!:$.
Correspondingly, a transformed functional is not well-defined in $\hat{\CA}$,
that is, it is divergent in terms of $\hat{\CA}$.
It should be a well-defined element of the new normal ordered module algebra $\hat{\CA}'$,
which is isomorphic to the twisted module algebra $\CA_{\CF'_0}$ with the product $*_{\CF'_0}$.
The relation between $\CA_{\CF'_0}$ and $\hat{\CA'}$ is the same as that 
discussed in \S \ref{sec:quantization}.
For example, the analogue of (\ref{coboundaryrelation}),
$\CF'_0= ({\CN'}_0^{-1} \otimes {\CN'}_0^{-1})\Delta (\CN'_0)$, holds.
We denote this new normal ordering with respect to $\CN'_0$ as 
$\,^\circ_\circ F \,^\circ_\circ $. 
Therefore, the VEV of the product of the vertex operators should also be 
defined through that of $\CA_{\CF'_0}$.
To see this, by rewriting the product (\ref{another viewpoint}), we obtain the 
relation 
between $\CA_{\CF'_0}$ and $\hat{\CA}'$ given by (\ref{product iso}) 
in \S \ref{sec:quantization} 
\bea
F[X'] *_{\CF'_0}G[X']
&=& {\CN'}_0^{-1} \act m \circ 
({\CN}'_0 \otimes {\CN}'_0) (u\otimes u) \act (F\otimes G) \nonumber \\
&=& {\CN'}_0^{-1} \act 
(\,^\circ_\circ F[X'] \,^\circ_\circ \,^\circ_\circ G[X'] \,^\circ_\circ)~.
\label{the other viewpoint}
\eea 
This and (\ref{change of normal ordered}) indicate that the VEV on $\hat{\CA}'$
should be the map 
$\langle \,\cdots \rangle'_0 =\tau' {\CN'}_0^{-1}: \hat{\CA}'\rightarrow \complex$.
For example, we have an $\CF'_0$-version of (\ref{F*G}) as
\be
\langle \,\,^\circ_\circ F[X'] \,^\circ_\circ \,^\circ_\circ G[X'] \,^\circ_\circ
\, \rangle'_0
= \tau' (F[X'] *_{\CF'_0}G[X'])~.
\label{VEV'}
\ee
Of course, it also coincides with the 
VEV without the prime through (\ref{trivial identity}).
Therefore, the transformation of the normal ordered module algebra under 
the change of 
variables requires the change of the normal ordered module algebra itself
to be consistent with that of the twisted module algebras.

There is also a direct correspondence between $\hat{\CA}'$ and $\hat{\CA}$.
For the product, we obtain from (\ref{change of normal ordered})
\bea
\,^\circ_\circ F[X'] \,^\circ_\circ \,^\circ_\circ G[X'] \,^\circ_\circ
&=& \left(\CN'_0 \act F[X']\right)\left(\CN'_0 \act G[X']\right) 
\nonumber\\
&=& m \circ (u\otimes u) ({\CN}_0 \otimes {\CN}_0)\act (F\otimes G) \nonumber \\
&=& u \act (:\!F[X]\!::\!G[X]\!:),
\label{product in A'}
\eea
which is also derived from (\ref{the other viewpoint}).
The corresponding VEVs on $\hat{\CA}'$ and $\hat{\CA}$ coincide
\be
\langle \,\,^\circ_\circ F[X'] \,^\circ_\circ \,^\circ_\circ G[X'] \,^\circ_\circ
\, \rangle'_0
=\langle \,:\!F[X]\!::\!G[X]\!:
\, \rangle_0~,
\label{VEV=VEV}
\ee
which follows from each definition of the VEV and (\ref{product in A'}).

\paragraph{}
The whole structure together with the isomorphism of twisted module algebras is as follows. 
We define a map $\hat{\rho}:\hat{\CA}\rightarrow \hat{\CA}'$
as $\hat{\rho} (:\!F[X]\!:)= u\,\act :\!F[X]\!:$.
Then (\ref{change of normal ordered}) is written as
$\CN'_0 \act \rho (F) = \hat{\rho} (\CN_0 \act F)$ and 
(\ref{product in A'}) is written as
$\CN'_0 \act \rho (F*_{\CF_0} G) = \hat{\rho} (\CN_0 \act (F*_{\CF_0} G))$.
This means that $\hat{\rho}$ is a (formal) algebra isomorphism, and 
the following diagram commutes,
\bea
\begin{array}{rcl}
\CA_{\CF_0} & \xrightarrow{\rho } &  \CA_{\CF'_0} \\
\CN_0\,\act  \downarrow & & \downarrow  \CN'_0\,\act \\
 \hat{\CA} & \xrightarrow{\hat{\rho}} & \hat{\CA'} 
\end{array}~.
\label{diagram2}
\eea
This shows the consistency of the isomorphism 
$\hat{\rho}$ between two normal ordered
module algebras,
but it is formal, that is, the relation is between two different 
divergent series.
Contrary to the case of twisted module algebras, changing the 
background requires the change of the normal ordering,
which corresponds to the new mode expansion in the operator formulation.

Note that in (\ref{product in A'}) $u$ acts as a group like element. 
This agrees with the transformation law inside the path integral, where
the variation is taken as if classical functionals are being 
considered, but each insertion is understood as being normal ordered.

Of course, 
if $\xi$ is in a twist-invariant Hopf subalgebra, such as $U(\CP)$ 
in our example,
the change of variables does not change the twisted 
$\CA_{\CF'_0}=\CA_{\CF_0}$ or the normal ordered $\hat{\CA'}=\hat{\CA}$ module algebras. 
In this case, the transformation is closed within $\hat{\CA}$ and 
has a well-defined 
meaning even in the operator formulation.

In the low-energy effective theory derived from the worldsheet theory, 
fields in space-time are associated with normal ordered vertex 
operators rather than with 
twisted module algebras.
If we consider the diffeomorphism beyond the Poincar\'e invariant theory,
we should be careful when considering 
the above change of the normal ordering. 
Even in such an application, the twist is the only simple way to
treat these changes systematically.
In any case, the space-time symmetry is governed by a
 single twist defining a background 
and its infinitesimal change under the diffeomorphism.

\subsection{Ward-like identities}
\label{sec:identities}

The difference between the path integral identities (\ref{general variation}) 
and the Hopf algebra counterpart 
(\ref{trivial identity}) or (\ref{VEV=VEV}) is 
the explicit appearance or absence of the action $S_0$.
The VEV in the Hopf algebra (\ref{def of true VEV}) is based on the 
twist element, 
while the action $S_0$ is not needed. In this sense in the Hopf 
algebra approach, the twist element has a more fundamental role
than the action.
However, it is also useful if we can 
compare it with the path integral 
expression, and we also obtain the relation to the action from the 
Hopf algebra viewpoint.

Actually, in the problem considered here, 
using the fact that we are dealing with a free theory, we can directly 
 derive identities concerning $S_0$ 
in terms of the Hopf algebra action.
Using them, we attempt to derive 
the same type of identity as (\ref{general variation}).

To this end, we start with the standard relation 
derived from (\ref{action}),
\be
{\delta S_0 \over \delta X^\mu (z)}
=-{1\over \pi\alpha'} \p\bar{\p} X_\mu (z)~.
\label{ppX}
\ee
Then we find an example of an identity:
The action $S_0$ itself satisfies
\bea
S_0 &=& \half\int \! d^2 z d^2 w \ \left({1\over \pi\alpha'} \p\bar{\p} X_\mu (z)\right)
G_0^{\mu\nu}(z,w)
 \left({1\over \pi\alpha'} \p\bar{\p} X_\nu (w)\right), \nonumber \\
&=& \half m \left( \int \! d^2 z d^2 w \ G_0^{\mu\nu}(z,w)
{\delta \over \delta X^\mu (z)}
\otimes {\delta \over \delta X^\nu (w)}\right) 
\act (S_0 \otimes S_0) , \nonumber \\
&=& -\half m \circ F_0 \act (S_0 \otimes S_0)~,
\label{S=FSS}
\eea
where in the first line 
$\p_z \bar{\p}_z G_0^{\mu\nu}(z,w)=-\pi\alpha' \eta^{\mu\nu} \delta^{(2)} (z-w)$
and integration by parts is used, 
in the second line \eqref{ppX} is inserted,
and in the last line $F_0$ is defined through $\CF_0 = e^{F_0}$.

In a similar manner, the functional derivative is also rewritten 
using \eqref{ppX} and $F_0$.
By a direct calculation, we find
\bea
{\delta \over \delta X^\mu (z)} \act I[X] 
&=& -m \circ F_0 \act \left(
{\delta S_0 \over \delta X^\mu (z)} \otimes I[X]
\right) \nonumber \\
&=& -m \circ \left({\delta \over \delta X^\mu (z)}\otimes 1 \right) 
F_0 \act \left(S_0 \otimes I[X] \right).
\label{dI=FSI}
\eea
Here in the second line we used the commutativity between 
the functional derivative and $F_0$. 
This is proved by noting that 
\bea
F_0 \act \left(S_0 \otimes I[X] \right)
=&& -{1\over \pi\alpha'}\int \!\! d^2 z \!\!\int \!\! d^2 w\, G_0^{\mu\nu} (z,w) 
\left( \p\bar{\p} X_\mu (z) \otimes 
{\delta I[X]  \over \delta X^\nu (w)} \right)\nonumber \\
=&& -{1\over \pi\alpha'} \int \!\! d^2 z \!\!\int \!\! d^2 w\, 
\p_z \bar{\p}_z G_0^{\mu\nu} (z,w)
X_\mu(z) \otimes
{\delta I[X]  \over \delta X^\tau (w)}
  \nonumber \\
=&&\int \!\! d^2 z X_\mu (z) \otimes {\delta I[X]  \over \delta X^\mu (z)},
\eea 
where integration by parts and the defining relation of the 
Green function are again used.
It is straightforward to extend to the action of a vector field 
$\xi \in {\mathfrak X}$ as
\be
\xi \act I[X] = -m \circ (\xi \otimes 1) F_0 
\act \left(S_0 \otimes I[X] \right).
\label{xI=xFSI}
\ee
Therefore, the action of any vector field on a functional 
can be rewritten using the action $S_0$.
From the Hopf algebra viewpoint, we do not start with the 
action $S_0$ but with 
the twist element $\CF_0$. 
In this context, the identities (\ref{xI=xFSI}) and (\ref{S=FSS}) are 
regarded as defining the action $S_0$ of the theory from the Hopf 
algebra action.
Note that they simply reflect the fact that the Green function is the 
inverse of the second-order 
differential operator
defining the equation of motion (\ref{ppX}).
Therefore, these relations and the following argument can be 
generalized to any theory
on the worldsheet with a quadratic action, and are not limited 
to (\ref{action}). 
This corresponds to any twist element of the type (\ref{F_0}) with an 
appropreate 
propagator.
Although we do not derive the boundary contribution explicitly, 
(\ref{xI=xFSI}) also holds for the worldsheet with boundaries.

One can also obtain a similar identity to (\ref{xI=xFSI}) 
with $\CF_0^{-1}$ instead of $F_0$.
Because $S_0$ is quadratic in $X$, we can explicitly calculate it as
\bea
m \circ (\xi \otimes 1) \CF_0^{-1} 
\act \left(S_0 \otimes I[X] \right) 
&=& m \circ (\xi \otimes 1) \left\{ 1-F_0+\half F_0^2 \right\} 
\act \left(S_0 \otimes I[X] \right), \nonumber \\
&=& (\xi \act S_0)I[X] + \xi \act I[X],
\label{xFSI=xSI+xI}
\eea
where in the second line \eqref{xI=xFSI} is used for the $F_0$ term.
The $F_0^2$ term is given by 
\begin{align}
F_0^2 \act \left(S_0 \otimes I[X] \right)
&= \int \!\! d^2 z \!\!\int \!\! d^2 w\, G_0^{\mu\nu} (z,w) 
\left(1 \otimes 
{\delta^2 I[X]  \over \delta X^\mu (z) \delta X^\nu (w)}\right) \nonumber \\
&= -2 \left(1 \otimes 
N_0 \act I[X] \right) ~.
\label{divergent term}
\end{align}
but it has no contribution since it vanishes when acting 
$\xi \otimes 1$ further on r.h.s., because $\xi \act 1=0$.
Note that (\ref{divergent term}) itself is 
divergent owing to the coincident point.
Applying the map $\tau$ to both sides of (\ref{xFSI=xSI+xI}), 
the first term in the r.h.s. of (\ref{xFSI=xSI+xI}) vanishes, 
and we have an identity for the VEV:
\be
\tau\circ m \circ (\xi \otimes 1) \CF_0^{-1} 
\act \left(S_0 \otimes I[X] \right) 
= \tau \left(\xi \act I[X] \right). 
\label{general identity for Hopf}
\ee
We can consider that the identity (\ref{general identity for Hopf})  
corresponds to (\ref{general variation}) in the path integral.
To see this more explicitly, let us rewrite the l.h.s. of (\ref{xFSI=xSI+xI}) as
\be
m \circ \CF_0^{-1} \left(\xi \act S_0 \otimes I[X] \right) 
+ m \circ \left[\xi \otimes 1, \CF_0^{-1} \right]
\act \left(S_0 \otimes I[X] \right)~.
\label{lhs}
\ee
Note that each term in (\ref{lhs}) is potentially divergent by the 
same reasoning as above,
thus care must be taken. 
Using the relations obtained in \S \ref{sec:quantization},
\bea
&&\tau \left(\xi \act I[X] \right) = \langle :\!\xi \act I[X]\!: \rangle_0 
= \langle ~\tilde{\xi}~ \act :\! I[X]\!: \rangle_0~,\nonumber \\
&&\tau \circ m \circ \CF_0^{-1} \left(\xi \act S_0 \otimes I[X] \right) 
=\tau \left((\xi \act S_0) *_{\CF_0} I[X] \right)
=\langle :\!(\xi \act S_0)\!: \ :\!I[X]\!: \rangle_0,
\eea
the identity (\ref{general identity for Hopf}) reduces to
\be
0= \langle ~\tilde{\xi}~ \act :\! I[X]\!: \rangle_0 
- \langle :\!(\xi \act S_0)\!: \ :\!I[X]\!: \rangle_0
-\tau\big( m \circ \left[\xi \otimes 1, \CF_0^{-1} \right]
\act \left(S_0 \otimes I[X] \right)\big) ~.
\label{general identity for Hopf2}
\ee
The identity (\ref{general identity for Hopf2}) is obtained by a simple 
rewriting of the Hopf algebra action, but remarkably, 
it appears to be similar to (\ref{general variation}). 
This also suggests that 
the last term can be identified with the variation of the measure 
$\CD X$, but we cannot conclude it at this stage.
Note also that $\hat{\CA'}$ is a suitable description of the change 
of variable 
as argued in \S \ref{sec:change of variables}, while here 
$\hat{\CA}$ is used.\footnote{
These two descriptions are related by the divergent series, 
which is related to the potential divergence noted above.}

Furthermore, (\ref{general identity for Hopf2}) contains the same 
type of information about identities 
(\ref{general variation}) derived in the 
path integral formalism as follows:
\begin{enumerate}
\item[(i)] If $\xi \in {\mathfrak C}$, then the last term vanishes since 
$\left[\xi \otimes 1, \CF_0^{-1} \right]=0$ 
and we obtain a Schwinger-Dyson type equation similar to (\ref{SDeq}).
\be
\tau(\xi\triangleright I[X])=\tau((\xi\triangleright S_0)*_{\CF_0} I[X])
\ee
or equivalently
\be
0= \langle ~\tilde{\xi}~ \act :\! I[X]\!: \rangle_0 
- \langle :\!(\xi \act S_0)\!: \ :\!I[X]\!: \rangle_0.
\ee
In this case, $\xi$ is not affected by the twist since
$\Delta_{\CF_0}(\xi)=\Delta (\xi)$, or equivalently, $\tilde{\xi}=\xi$.
Therefore, the action of $\xi$ here satisfies the Leibniz rule
(we do not need to consider the difference between $\hat{\CA}$ and $\hat{\CA'}$).
\item[(ii)] If $\xi$ is a classical symmetry of the theory, we have $\xi \act S_0 =0$.
Moreover, if the last term vanishes, then we obtain the Ward identity
\be
0=\tau(\xi\triangleright I[X])= \langle ~\tilde{\xi}~ \act :\! I[X]\!: \rangle_0 .
\ee
With the same reasoning as above, 
$\xi$ is still primitive under the twist so that
the action of $\xi$ splits into the sum of the transformations for 
each local functional contained in 
the functional $I[X]$.
\item[(iii)]  As in the path integral case, considering the 
variation $\rho\xi$ 
instead of $\xi$ in the above derivation, 
we obtain the relation including the Noether current.
\item[(iv)] For general $\xi$, there are contributions from the variation of 
the action $S_0$ as well as the last term, and the variation of the insertion $I[X]$
is not split into individual variations.
Nevertheless, it is compactly written as the twisted Hopf algebra action $\xi \act I[X]$.
\end{enumerate}
We do not derive the path integral identities (\ref{general variation}) 
using the change of variables argument in \S 
\ref{sec:change of variables}.
However, as we have seen above, the symmetry is characterized again 
as the twist-invariant Hopf subalgebra 
of $\CH$ which keeps $S_0$ invariant.
In this case, the transformation law of the twisted 
Hopf subalgebra is given in the same form as the classical 
transformation. 
In our model, this subalgebra is the universal enveloping algebra of the 
Poincar\'e-Lie algebra $U(\CP)$, as already remarked.
The action of $u\in U(\CP)$ keeps $S_0$ invariant, and thus the 
quantized transformation is the same as the classical transformation.
It also satisfies the Leibniz rule and leads to 
the ordinary Ward identity 
of the form (\ref{Ward id}).

\section{Conclusion and discussion}

We have investigated the Hopf algebra structure in the quantization of
the string worldsheet theory in the target space $\real^d$.
It gives a unified description of both 
the quantization and the space-time symmetry
simply as a twist of the Hopf algebra.

In the functional description of the string, 
we found that the module algebra $\CA$ 
of classical functionals as well as 
the Hopf algebra $\CH$ of functional derivatives 
correspond to space-time diffeomorphisms and worldsheet variations.
They are background-metric-independent in nature,
but the choice of a twist element $\CF_0$ fixes the background.
Twisting them by $\CF_0$ gives the covariant quantization on this background.
We have seen that the twist is formally trivial and that it also 
characterizes the normal ordering.
Therefore, the twist is equivalent to the description in the path integral as well as the 
operator formulations.
On the other hand, the twist also characterizes the broken and unbroken 
space-time symmetry.
In our fixed Minkowski background, 
the symmetry of the Poincar\'e transformations 
remains unbroken as a twist-invariant Hopf subalgebra $U(\CP)$.
The remaining transformations, which are 
broken in the Minkowski background,
are still retained as a twisted Hopf algebra.
We have explicitly seen that the classical diffeomorphism 
in $\real^d$ is realized 
as a twisted diffeomorphism in such a way that the background $\eta_{\mu\nu}$ is fixed.
Therefore, it is a good starting point for discussing 
the background independence 
in full generality.

We give an outlook regarding this work.
Our consideration is limited to the worldsheet theory of strings, 
but it is also important to relate it to the low-energy effective theory
including gravity, where there is the classical general covariance.
For that purpose, we have to further investigate 
particular correlation functions and the S-matrix.
Note that this is merely the on-shell equivalence and 
there is always a difficulty of field redefinition ambiguities.

Another issue that we did not treat 
in this paper is the local symmetries on the worldsheet,
in particular the conformal symmetry.
It restricts the possible backgrounds and it is also 
necessary to obtain the spectrum of the theory.
At the level of our treatment in this paper, 
the conformal symmetry should be additionally imposed.
However, it would be possible to incorporate it by enlargement of 
 the Hopf algebra,
which probably touches upon the Hopf algebraic 
structure in conformal field theory 
\cite{ChariPressley}.

Our scheme of quantization with normal ordering
works at least for any twist $\CF_0 \in U({\mathfrak C})\otimes U({\mathfrak C})$
given by a Green function, corresponding to free theories. 
In this context, a background with a non-zero $B$-field can be 
considered in the same manner. 
This will be discussed in Ref.\citen{inpreparation}.
On the other hand, a twist element can be any element in $\CH\otimes \CH$ satisfying
 the cocycle condition and counital condition.  
Thus, the twist element is not necessarily given by a Green function. 
Such a nonabelian twist would correspond to the interacting theory 
on the worldsheet.
We would also like to consider target spaces other than $\real^d$. 
In that case, the general strategy proposed in this paper, 
the unified treatment of the worldsheet and target space variations as a
Hopf algebra, is also expected to work.
This would shed new light 
on the understanding of the quantization of strings in more general 
backgrounds and also 
on the geometric structure of strings as quantum gravity.

\section*{Acknowledgements}
The authors would like to thank
Y.~Shibusa, K.~Ohta, H.~Ishikawa and U. Carow-Watamura
for useful discussions and valuable comments. 
This work is supported in part by the Nishina Memorial Foundation (T.A.)
and by a Grant-in-Aid for Scientific Research from the Ministry of Education, Culture, Sports, Science and Technology, Japan, No. 19540257(S.W.).

\appendix

\section{Hopf Algebra}
\label{Hopf}
Here we introduce some definitions and our conventions regarding the 
Hopf algebra and its action.

A Hopf algebra $(H;\mu,\iota,\Delta,\epsilon,S)$ over a field $k$  
is a $k$-vector space $H$ equipped with the following linear maps
\begin{eqnarray} 
&& \mu: H \otimes H \rightarrow H~~(\text{multiplication}), 
~~~~\iota: k \rightarrow H~~(\text{unit}),    \\
&&   \Delta: H \rightarrow H \otimes H~~(\text{coproduct}),~~~ 
\epsilon: H \rightarrow k~~(\text{counit}), \\ 
&&S:H \rightarrow H~~(\text{antipode}),
\end{eqnarray}
(We also denote $\mu(h\otimes g)=hg$ and $\iota(k) =k 1_H$.)
and satisfying the following relations: 
\begin{eqnarray} 
&& (fg)h=f(gh), 
~~~~h1_H =h=1_H h, \nonumber \\ 
&&(\Delta \otimes {\rm id})\circ\Delta (h)= ({\rm id} \otimes \Delta )\circ\Delta (h), 
~~~~(\epsilon \otimes {\rm id})\circ \Delta (h)=h=({\rm id}\otimes \epsilon )\circ \Delta (h)
\nonumber  \\ 
&&  \mu\circ(S \otimes {\rm id})\circ \Delta(h)=
\epsilon(h)=\mu\circ({\rm id} \otimes S)\circ\Delta (h) \nonumber\\
&&\Delta(gh)=\Delta(g)\Delta(h),~~~\epsilon(gh)=\epsilon(g)\epsilon(h),~~~
S(gh)=S(h)S(g)
\end{eqnarray} 
for ${}^\forall f,g,h \in H$.
An universal enveloping algebra $U({\mathfrak g})$ of a Lie algebra 
${\mathfrak g}$ is a Hopf algebra: 
it is a tensor algebra 
generated by the elements of ${\mathfrak g}$ 
modulo the Lie algebra relation.     
The remaining maps are defined for $g \in {\mathfrak g}$ by 
\begin{eqnarray} 
&&\Delta(g)=g \otimes 1 +1 \otimes g,~~~
\epsilon(g)=0,~~~S(g)=-g, 
\end{eqnarray} 
and are extended for arbitrary elements $u \in U({\mathfrak g})$ 
by the (anti)homomorphism property of $\Delta$ and $\epsilon$ ($S$).

A (left) $H$-module $A$ of a Hopf algebra $H$ is a module of $H$ 
that is an algebra, i.e.,
a $k$-vector space 
equipped with a map $\alpha :H \otimes A\rightarrow A$ called an $H$-action 
(we denote it as $\alpha (h\otimes a)=h\act a$) such that 
\begin{eqnarray} 
(gh) \act a=g \act( h\act a),~~~~1_H \act a=a.
\end{eqnarray} 
If in addition $A$ is a unital algebra with a multiplication 
$m:A\otimes A \rightarrow A$ such that 
\begin{eqnarray} 
h \act m(a\otimes b)=m\circ \Delta(h)\act (a\otimes b),~~~~h\act 1=\epsilon (h)1,
\label{MA-condition}
\end{eqnarray}
then $A=(A\,;m)$ is called an $H$-module algebra.

The Drinfeld twist of a Hopf algebra $H$ is given by an invertible element 
$\CF\in H \otimes H$ such that \cite{Drinfeld}
\begin{eqnarray}
&&{\cal F}_{12}(\Delta \otimes {\rm id}){\cal F}={\cal F}_{23}
({\rm id} \otimes \Delta){\cal F},~~(\text{cocycle condition}) \nonumber \\ &&
({\rm id} \otimes \epsilon){\cal F}=1=(\epsilon \otimes {\rm id}){\cal F},
~~(\text{counital condition})
\end{eqnarray}
where the suffix of ${\cal F}_{12}$ denotes that 
${\cal F}$ acts on the first and second 
elements of $H\otimes H \otimes H$.
Then a twisted Hopf algebra 
$(H;\mu,\iota,\Delta_{\cal F},\epsilon,S_{\cal F})$, 
defined by
\begin{eqnarray}
&&\Delta_{\cal F}(h)={\cal F}\Delta (h) {\cal F}^{-1}, \nonumber \\ &&
S_{\cal F}(h)=U S(h) U^{-1}, ~~\text{where}~~
U := \mu\circ({\rm id} \otimes S)({\cal F})
\end{eqnarray}
for ${}^{\forall} h \in H$,
satisfies all the axioms. We denote this as $H_{\CF}$.

For an $H$-module algebra $A=(A\,;m)$,
there is an associated $H_{\CF}$-module algebra $A_{\CF}=(A;m_{\cal F})$
with a twisted multiplication $m_{\cal F} :A_{\CF}\otimes A_{\CF}\rightarrow A_{\CF} $,
which is also denoted as $\ast_{\CF}$. 
For $a,b \in A$ it is given by
\begin{eqnarray}
a*_{\CF} b= m_{\cal F}(a \otimes b):=m \circ \CF^{-1} \act (a\otimes b)
\end{eqnarray}
and is associative owing to the cocycle condititon.
Condition (\ref{MA-condition}) is proved as  
\begin{eqnarray}
h \act  m_{\cal F} (a \otimes b) 
=m \circ \Delta(h) {\cal F}^{-1} \act (a \otimes b)
=m_{\cal F} \circ \Delta_{\cal F}(h) \act(a \otimes b).
\end{eqnarray}

\section{Proof of the Cocycle Condition}
\label{sec:cocycle-cond}

Here, we show that an element in $U({\mathfrak C})\otimes U({\mathfrak C})
\subset \CH\otimes \CH$ of the form
\begin{equation}
\CF=\exp\left(-{\int \!d^2z \!\!\int \!d^2w \,
G^{\mu\nu}(z,w)
\frac{\delta}{\delta X^{\mu}(z)}
\otimes\frac{\delta}{\delta X^{\nu}(w)}}\right),
\label{generaltwist}
\end{equation}
is a twist element and can be used to obtain the twist Hopf algebra $\CH_{\CF}$.
It is clearly invertible and counital, $({\rm id} \otimes \epsilon)\CF=1$.
The 2-cocycle condition,
$
\CF_{12}(\Delta \otimes {\rm id})\CF=\CF_{23}
({\rm id} \otimes \Delta)\CF
$,
is satisfied because 
the two sides can be written as
\be
\begin{cases}
\CF_{12}(\Delta \otimes {\rm id})\CF &=
\CF_{12}e^{-\int d^2z d^2w \,\,
G^{\mu\nu}(z,w)
\left\{1 \otimes\frac{\delta}{\delta X^{\mu}(z)}
+\frac{\delta}{\delta X^{\mu}(z)}\otimes 1 \right\}
\otimes\frac{\delta}{\delta X^{\nu}(w)}} \\
&=e^{-\int d^2z d^2w \,\,
G^{\mu\nu}(z,w)
\left\{\frac{\delta}{\delta X^{\mu}(z)}
\otimes\frac{\delta}{\delta X^{\nu}(w)}\otimes 1 +
1 \otimes\frac{\delta}{\delta X^{\mu}(z)}
\otimes\frac{\delta}{\delta X^{\nu}(w)}
+\frac{\delta}{\delta X^{\mu}(z)}\otimes 1
\otimes\frac{\delta}{\delta X^{\nu}(w)}\right\}} \\
\CF_{23}({\rm id} \otimes \Delta)\CF 
& =e^{-\int d^2z d^2w \,\,
G^{\mu\nu}(z,w)
\left\{
1 \otimes
\frac{\delta}{\delta X^{\mu}(z)}\otimes \frac{\delta}{\delta X^{\nu}(w)}
+\frac{\delta}{\delta X^{\mu}(z)}
\otimes 1 \otimes\frac{\delta}{\delta X^{\nu}(w)}
+\frac{\delta}{\delta X^{\mu}(z)}\otimes \frac{\delta}{\delta
X^{\nu}(w)} \otimes 1 
\right\}},
\end{cases}
\nonumber
\ee 
where we used $\Delta({\delta \over \delta X})=1\otimes {\delta \over \delta X}
+{\delta \over \delta X}\otimes 1$ and the fact that $\Delta$ is an algebra homomorphism.

Next let us assume that the ``propagator'' in the exponent is symmetric,
$G^{\mu\nu}(z,w)=G^{\nu\mu}(w,z)$. 
Then, from the same argument as that in \S \ref{sec:quantization}, 
the twist element $\CF$ is a coboundary, $\CF=(\CN^{-1} \otimes \CN^{-1})\Delta (\CN)$.
On the other hand, the antipode should be twisted, 
$S\ne S_{\CF}=US U^{-1}$, in general.
In fact, $U$ is not $1$ and is given explicitly by
\begin{eqnarray}
U&=&\mu\circ({\rm id}\otimes S) \CF 
=e^{\int d^2z \int d^2w \,\,
G^{\mu\nu}(z,w)
\frac{\delta}{\delta X^{\mu}(z)}
\frac{\delta}{\delta X^{\nu}(w)}}
=\CN^{-2},
\end{eqnarray}
where we used $S({\delta \over \delta X})=-{\delta \over \delta X}$ 
and the fact that 
$S$ is an algebra antihomomorphism.

\section{Trivial Twists}
\label{trivial}
In this appendix, we consider the case when a Hopf algebra twist is trivial 
in the cohomologous sense as discussed in Ref.\citen{Majid}.
Let $H$ be a Hopf algebra.
For any invertible element $\gamma \in H$ s.t. $\epsilon \gamma =1$,
the corresponding element $\p \gamma \in H\otimes H$
\bea
\p \gamma =(1\otimes \gamma )(\gamma \otimes 1)\Delta \gamma^{-1} 
=(\gamma \otimes \gamma )\Delta \gamma^{-1}
\eea
is a counital 2-cocycle, where $\p$ is defined in Ref.\citen{Majid}. 
However, since $\p\p\gamma =0$, it is a trivial 2-cocycle (called a coboundary) .
More generally, two 2-cocycles $\psi,\chi$ are said to be cohomologous if they 
are related by a coboundary $\gamma$ as
\bea
\psi = (\gamma \otimes \gamma )\chi \Delta \gamma^{-1}.
\eea
Then, it is shown that two Hopf algebras $H_\chi $ and $H_\psi $ 
twisted by $\chi,\psi$ are isomorphic as Hopf algebras under an inner automorphism. 
This isomorphism is given by
$\pi: H_\psi \rightarrow H_\chi : h \mapsto \pi(h)=\gamma^{-1} h \gamma$.
Here, the coproduct in $H_\psi $ can be written 
for ${}^{\forall} h \in H_\psi $ as
\bea
\label{cohomologous coproduct}
\Delta_\psi (h) &=& \psi (\Delta h) \psi^{-1} 
=(\gamma \otimes \gamma )\chi (\Delta \gamma^{-1})(\Delta h)(\Delta \gamma)
\chi^{-1} (\gamma^{-1} \otimes \gamma^{-1} ) \nonumber \\
&=& (\gamma \otimes \gamma )(\Delta_\chi(\gamma^{-1} h \gamma))
 (\gamma^{-1} \otimes \gamma^{-1} ).
\eea
Because $(\pi\otimes \pi) (h_1 \otimes h_2) \mapsto 
(\gamma^{-1} \otimes \gamma^{-1})(h_1 \otimes h_2)(\gamma \otimes \gamma)$,
it implies the coalgebra isomorphism 
$(\pi\otimes \pi) (\Delta_\psi (h)) = \Delta_\chi (\pi(h)) $.
The other structures are also easily verified to be isomorphic.

In the same way, if an $H_\chi$-module algebra $A_\chi$ and an 
$H_\psi$-module algebra $A_\psi$ are obtained by twisting the same 
$H$-module algebra $A$, then they are isomorphic as module algebras.
Let us define 
$\tilde \pi: A_\psi \rightarrow A_\chi : f \mapsto \gamma^{-1} \triangleright f$.
Then a module map from 
the Hopf algebra action $H_\psi \triangleright A_\psi$ to 
that for $H_\chi \triangleright A_\chi$ is given as
\bea
\label{module iso}
\tilde \pi(h \,\triangleright f) 
&=&\gamma^{-1} \triangleright (h \,\triangleright f) 
=(\gamma^{-1} h \gamma)\triangleright (\gamma^{-1} f) \nonumber \\
&=&\pi(h) \triangleright \tilde \pi (f).
\eea
$\tilde \pi$  also relates the twisted products $m_\psi$ and  $m_\chi$ as
\bea
\tilde \pi (m_\psi (f\otimes g)) 
&=&\gamma^{-1} \,\triangleright m \, (\psi^{-1} \triangleright (f\otimes g)) 
\nonumber \\
&=&m \,(\Delta \gamma^{-1})(\Delta \gamma)\chi^{-1} (\gamma^{-1} 
\otimes \gamma^{-1} )
\triangleright (f\otimes g) \nonumber \\
&=& m_\chi
(\tilde \pi (f)\otimes \tilde \pi(g)). 
\eea

In particular, if $\chi =1\otimes 1$, the Hopf algebra $H_\psi $
twisted by the coboundary $\psi =\p \gamma$ is isomorphic to the original 
Hopf algebra $H$, that is, 
the twisting is undone by the inner automorphism $\pi: H_\psi \rightarrow H$.

\section{Poincare Symmetry and $\CF_0$}
\label{sec:Poincare}

Here we prove that $U(\CP)$ is the invariant Hopf subalgebra 
under the twist $\CF_0$ (\ref{F_0}).
For this, it is sufficient to show that none of the coproducts of 
the generators of the Poincare-Lie algebra $\CP$
are modified.
Recalling that $\CF_0=e^{F_0}$, we must show that the
coproduct of the generators
$P_\mu$ and $L_{\mu\nu}$ in (\ref{generators}) 
commutes with $F_0$ in (\ref{F_0}). 
Since $P_\mu \in {\mathfrak C}$, 
it is apparent that $\Delta(P_\mu)=P_\mu\otimes 1+1\otimes P_\mu$ 
commutes with $F_0$. 
For the Lorentz generators, by using the fact that the propagator is of the form 
$G_0^{\mu\nu}(z,w)=\eta^{\mu\nu} G_0(z,w)$, we can verify this as
\bea
&&\left[F_0, \Delta(\epsilon_{\mu\nu}L^{\mu\nu}) \right] 
=\left[ F_0,~( 
\epsilon_{\mu\nu}L^{\mu\nu}\otimes 1+1\otimes \epsilon_{\mu\nu}L^{\mu\nu}) \right]\cr
&&=-\int \!d^2z \!\!\int \!d^2w \, G_0^{\rho\lambda}(z,w) \left(
\epsilon_{\rho\nu}{\delta\over\delta X_\nu (z)}\otimes {\delta\over\delta X^\lambda (w)}
+{\delta\over\delta X^\rho (z)}\otimes \epsilon_{\lambda\nu}{\delta\over\delta X_\nu (w)}
\right)\cr
&&=- \int \!d^2z \!\!\int \!d^2w \, G_0(z,w)
(\epsilon^{\lambda\nu}+\epsilon^{\nu\lambda})
{\delta\over\delta X^\nu (z)}\otimes {\delta\over\delta X^\lambda (w)} =0
\eea
where we have used this fact and $G_0(z,w)=G_0(w,z)$ in the last step.


%


\begin{thebibliography}{99}

 
\bibitem{Polchinski}{J. Polchinski, {\it String Theory} (Cambridge 
Univ. Press, 1998.)}

\bibitem{Majid}{S. Majid, {\it Foundations of Quantum Group Theory} (Cambridge 
Univ. Press, 1995.)}

\bibitem{Oeckel}{S.~Majid and R.~Oeckl, Commun. Math. Phys. 
{\bf 205} (1999), 617; {{\tt math.QA/9811054}}.
}

\bibitem{Sitarz0102074}{
A.~Sitarz, Lett. Math. Phys. {\bf 58} (2001), 69; {{\tt math.QA/0102074}}.
}

\bibitem{Watts}{
P.~Watts, {{\tt hep-th/0003234}}.

\bibitem{Oeckl2}{
R.~Oeckl, Nucl. Phys. B {\bf 581} (2000), 559; {{\tt hep-th/0003018}}.}

\bibitem{KulishNishijima}{
M.~Chaichian, P.~P. Kulish, K.~Nishijima and A.~Tureanu, Phys. Lett. B 
{\bf 604} (2004), 98; {{\tt hep-th/0408069}}.}

\bibitem{Wess}{
P.~Aschieri, C.~Blohmann, M.~Dimitrijevic, F.~Meyer, P.~Schupp and J.~Wess, 
Class. Quantum Grav. {\bf 22} (2005), 3511; {{\tt hep-th/0504183}}.
P.~Aschieri, M.~Dimitrijevic, F.~Meyer and J.~Wess, Class. Quant. Grav. {\bf 23} (2006), 1883;  {{\tt hep-th/0510059}}.
}
\bibitem{Aschieri}{P.~Aschieri, F.~Lizzi and P.~Vitale, 
{\tt arXiv:0708.3002v2[hep-th]}.}



\bibitem{Kobayashi}{
Y.~Kobayashi and S.~Sasaki, Int. J. Mod.  Phys. A {\bf 20} (2005), 7175; 
{{\tt hep-th/0410164}}.}

\bibitem{Drinfeld}{
V. G. Drinfeld, Leningrad Math. J. {\bf 1} (1990), 1419.
}

\bibitem{SeibergWitten}{N.~Seiberg and E.~Witten, JHEP {\bf 09} (1999), 032; 
{{\tt hep-th/9908142}}.}

\bibitem{inpreparation}{In preparation.}

\bibitem{BFFLS}
F.~Bayen, M.~Flato, C.~Fronsdal, A.~Lichnerowicz and D.~Sternheimer, 
Ann. Phys. {\bf 111} (1978) 61;
F.~Bayen, M.~Flato, C.~Fronsdal, A.~Lichnerowicz and D.~Sternheimer, Ann.
  Phys. {\bf 111} (1978), 111.
}

\bibitem{Dito}{
J.~Dito,  Letters in Math. Phys. {\bf 20} (1990),  125. 
J.~Dito, Letters in Math. Phys. {\bf 27} (1993), 73.
}

\bibitem{GarciaCompean}{
H.~Garcia-Compean, J.~F. Plebanski, M.~Przanowski and F.~J. Turrubiates, 
J. Phys. A {\bf 33} (2000), 7935; {{\tt hep-th/0002212}}.
}

\bibitem{BrouderOeckl}{
C.~Brouder and R.~Oeckl, {{\tt hep-th/0208118}}.
}

\bibitem{BFO}{
C.~Brouder, B.~Fauser, A.~Frabetti and R.~Oeckl, J. Phys. A {\bf 37} (2004) 
5895; {{\tt hep-th/0311253}}.
}

\bibitem{AG}
L.~Alvarez-Gaume, F.~Meyer and M.~A. Vazquez-Mozo, Nucl. Phys. B {\bf 753} 
(2006), 92; {{\tt hep-th/0605113}}.

\bibitem{ChariPressley}{
V.~Chari, A.~ Pressley, {\it A Guide to Quantum Groups} (Cambridge Univ. 
Press), 1994.
}



\end{thebibliography}
\end{document}